\documentclass[12pt]{article}

\usepackage{times}
\usepackage{graphicx}
\usepackage{color}
\usepackage{ccaption}
\usepackage{verbatim}
\usepackage{mathtools}
\usepackage{amssymb,amsmath}
\usepackage{multirow}
\usepackage[table,xcdraw]{xcolor}
\usepackage{hhline}
\usepackage{setspace}
\usepackage{versions}
\usepackage{xr}
\usepackage{hyperref}
\usepackage{bm}

\usepackage[utf8]{inputenc} 
\usepackage{cite}
\usepackage{textcomp}
\usepackage{float}
\usepackage{longtable}
\usepackage{tikz}
\usepackage{placeins}
\usepackage{lipsum}

\usepackage{siunitx}
\DeclareSIUnit{\inch}{in}
\DeclareSIUnit{\liter}{L}

\usepackage{caption}
\usepackage{subcaption}

\usepackage[labelfont=bf]{caption}

\usepackage [autostyle, english = american]{csquotes}
\MakeOuterQuote{"}
\usepackage{setspace}
\newenvironment{sciabstract}{%
\begin{quote} \bf}
{\end{quote}}

\title{Ecological memory of hydrodynamic cues shapes growth and migration of motile microorganisms}

\author
{Narges Kakavand$^{1}$ and Anupam Sengupta$^{1,2^\ast}$
\\
\\
\normalsize{$^{1}$Physics of Living Matter, Department of Physics and Materials Science, }\\
\normalsize{University of Luxembourg, 162 A, Avenue de la Fa\"{i}encerie, L-1511, Luxembourg}\\
\normalsize{$^{2}$ Institute for Advanced Studies, University of Luxembourg, 2, Avenue de l’Université,}\\
\normalsize{L-4365, Esch-sur-Alzette, Luxembourg}\\
\normalsize{Email: $^\ast$anupam.sengupta@uni.lu}
}

\date{}
\includeversion{v1}
\excludeversion{v2}

\begin{document} 



\maketitle


\begin{sciabstract}

Microorganisms live in inherently dynamic environments where repeated fluctuations in biotic and abiotic factors shape their behaviour, physiology, and overall fitness. The concept of ecological memory—the lasting imprint of prior environmental conditions—suggests that past exposures can exert prolonged effects on microbial growth, stress resilience, and phenotypic expressions. For motile microorganisms inhabiting aquatic ecosystems, environmental variability is strongly mediated by fluid motion. Cells experience spatially and temporally heterogeneous flow fields, ranging from steady shear to intermittent turbulence. These hydrodynamic signals may generate a form of “hydrodynamic memory,” whereby prior exposure to specific flow regimes influences future growth and migratory behaviour. Yet, the emergence of such flow-induced memory, or its long-term consequences for trait evolution and population dynamics, remain unexplored. Here, we integrate millifluidic flow control, high-resolution cell tracking, and growth assays under tunable hydrodynamic cues to quantify growth and swimming behavior of \textit{Heterosigma akashiwo}, a model marine raphidophyte, across growth stages. Using two complementary perturbation scenarios, we investigate how the temporal structure of forcing shapes cellular responses across generations. In a \textit{standard scenario}, hydrodynamic cues follow static conditions, while in a \textit{reverse scenario}, flow exposure precede static growth. This design enables us to disentangle the effects of exposure history as well as its duration, and to evaluate how prior flow conditions modulate cellular sensitivity and generate legacy effects. Repeated hydrodynamic exposure produces measurable changes in doubling time and carrying capacity, as well as in gravitactic stability and swimming speed distributions, relative to populations maintained under static conditions. These history-dependent modifications alter growth phase progression and tolerance to subsequent flow perturbations. Our results establish a mechanistic framework for flow-induced memory in swimming microorganisms and demonstrate how prior fluid environments shape future physiological and migratory responses. This work advances predictive understanding of motile microbes in natural and engineered systems, where shifts in hydrodynamic regimes are an increasingly prominent feature of global environmental change.

\end{sciabstract}


\section*{Author Contributions}
Conceptualization, planning, administration, and supervision: A.S. Methodology: N.K. and A.S. Investigation, data and statistical analysis: N.K. with inputs from A.S. Writing, review \& editing: N.K. and A.S.

\section*{Data Availability Statement}
All relevant data supporting the findings are included in the manuscript. Additional datasets and analysis code (MATLAB scripts for image analysis and growth-curve fitting) are available from the corresponding author upon reasonable request.

\section*{Conflict of Interest Statement}
The authors declare that the research was conducted in the absence of any commercial or financial relationships that could be construed as a potential conflict of interest.

\section*{Acknowledgements \& Funding}
This work was carried out with the support of the FNR-Luxembourg's PRIDE Doctoral Training Unit ACTIVE (Active Phenomena Across Scales in Biological Systems, \url{https://cls.uni.lu/doctoral-training/}), with additional funding from the University of Luxembourg and the Luxembourg National Research Fund's ATTRACT Investigator Grant (Grant no. A17/MS/ 11572821/MBRACE to AS) as well as the CORE Grant (Grant no. C19/MS/13719464/ TOPOFLUME/Sengupta to AS). AS gratefully acknowledges the AUDACITY Grant (AUDACITY Grant no.: IAS-20/CAMEOS) from the Institute for Advanced Studies, University of Luxembourg for supporting this work.

\newpage

\maketitle

\section*{Introduction}

Microorganisms inhabit environments that are inherently dynamic, characterized by fluctuations in nutrient availability, temperature, pH, moisture and biotic interactions. Unlike short-lived, single-condition cultures, microbial populations in natural and host-associated ecosystems often experience repeated cues over sustained durations. There is growing evidence that such historical exposure can leave a lasting imprint on the physiology, growth dynamics and phenotypic traits of microorganisms—a phenomenon broadly referred to as ecological memory \cite{Vaas2017, Canarini2021, Liang2025}. In microbial systems, ecological memory manifests through history-dependent changes in growth, metabolism, motility, and stress tolerance, at both the cellular and population levels. Individual cells may retain imprints of past environments through transcriptional changes, protein inheritance or epigenetic modifications that influence future growth and metabolism under recurring conditions \cite{Vermeersch2022}. At the community level, ecological memory can be encoded in population structure and environmental modification: changes in community composition or in the chemical milieu created by microbes can bias responses to future exposures. For example, repeated exposure to specific nutrients in the human gut microbiome rapidly alters the abundance and metabolic activity of key taxa, enhancing metabolic potential upon subsequent encounters, a hallmark of microbial ecological memory \cite{Letourneau2022}. Moreover, ecological memory may not be confined to individual species but emerges from collective interactions, where environmental modification by one population creates feedback on the entire community’s growth trajectories and phenotypic traits \cite{Gajrani2025}. These processes contribute to phenotypic plasticity, niche construction, and altered stress responses that have significant implications not only for fundamental microbial ecology but also for applications such as bioremediation, pathogen control, and human health. Understanding the mechanisms, time scales and cross-generational consequences of microbial ecological memory is therefore critical for predicting and managing community interactions and feedback in a changing world.

In aquatic systems, where hydrodynamic variability is intrinsic, fluid flow represents a primary driver of such history-dependent responses. For instance, exposure to sustained shear or turbulence can alter flagellar synthesis, swimming speed distributions, chemotactic responsiveness, and the regulation of attachment-related pathways in bacteria \cite{Wolf2008, Gokhale2021, Rene2024}. These changes may persist beyond the initial perturbation, modifying lag times, growth rates, and resource acquisition efficiency upon subsequent exposure to similar or contrasting flow conditions. Governed by fluid motion across a wide range of spatial and temporal scales, microbes are exposed to microscale shear as well as turbulent mixing in lakes and oceans \cite{sengupta2023planktonic}. For swimming microorganisms such as motile bacteria and microalgae, fluid flow is not merely a background condition but a pervasive and information-rich environmental cue. Hydrodynamic forces structure nutrient gradients, regulate encounter rates, influence light exposure, and modulate mechanical stresses experienced by cells. Increasing evidence suggests that repeated or prolonged exposure to particular flows can have long-term effects on microbial physiology and behavior, driving a form of ecological memory that shapes subsequent growth and phenotypic expressions. 

In motile microalgae, hydrodynamic cues interact with gravitaxis, phototaxis, and gyrotactic stability, influencing vertical positioning and light harvesting. Repeated exposure to turbulent mixing can select for altered swimming behaviors, changes in cell morphology, and modifications of flagellar coordination. Such phenotypic adjustments may affect cell cycle progression and photosynthetic performance, thereby coupling mechanical history to population growth dynamics \cite{Sengupta2017Phytoplankton, Carrara2021Bistability}. The persistence and reversibility of these traits suggest that aquatic microorganisms may encode aspects of past flow environments through regulatory network reconfiguration, structural remodeling, or selection within phenotypically heterogeneous populations. Recurrent hydrodynamic regimes may bias population composition and trait distributions, encoding memory not only in individual cells but in the spatial organization and collective dynamics \cite{Pfreundt2023, Volpe2023}. In stratified lakes and estuaries, where mixing intensity fluctuates seasonally or episodically due to biotic and abiotic factors, such memory effects may contribute to recurring patterns of bloom formation, microbial succession, and biogenic mixing \cite{Sommer2017, Dinezio2023}. Understanding how fluid mechanical histories shape microbial growth and phenotype is increasingly important in the context of global change. Alterations in wind forcing, stratification strength, and extreme weather events are modifying hydrodynamic regimes across aquatic ecosystems. If swimming microorganisms retain ecological memory of prior flow conditions, their responses to these shifts may depend strongly on past exposure, challenging predictions based solely on present conditions. 

\begin{figure}
\centering
\includegraphics[width= \columnwidth]{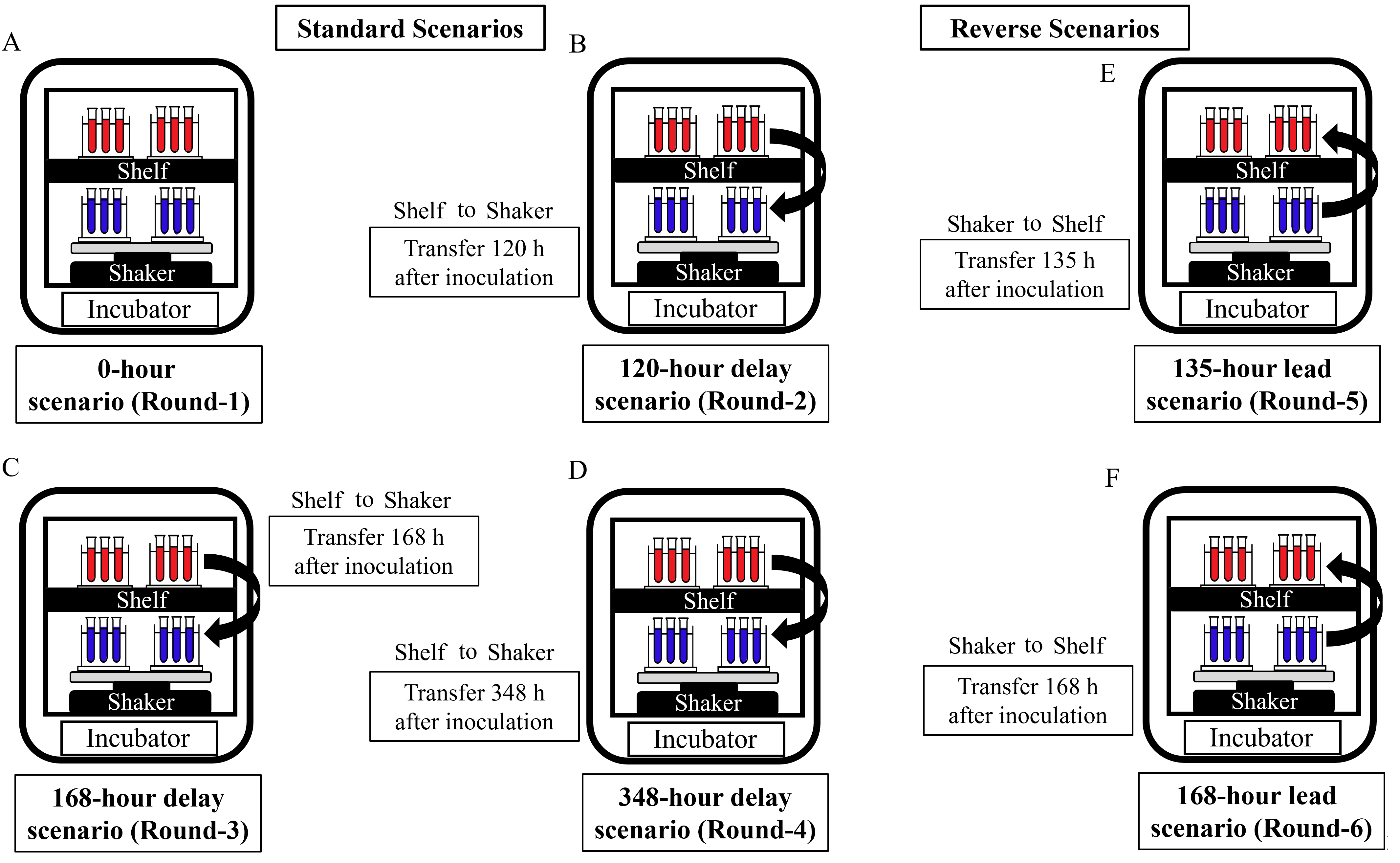}
\caption[Hydrodynamic scenarios]{\textbf{Tracking ecological memory in motile microorganisms.} 
(A)--(D): \textit{Standard} scenario (static $\rightarrow$ shaker): microbial growth initially under static condition (red hue), followed by hydrodynamic perturbations (blue hue).
(A) 0-hour: cells transferred to the shaker immediately post-inoculation, while control cultures are maintained under static conditions within the same incubator. (B) 120-hour \textit{delay}: 120 h (early exponential phase) of static condition, followed by hydrodynamic perturbation. Similarly, (C) captures 168 h delay, corresponding to mid-exponential phase. (D) 348 h delay: cultures are exposed to perturbation after 348 h of growth (stationary phase) under static condition. (E)--(F): \textit{Reverse} scenarios (shaker $\rightarrow$ static).
(E) 135 h \textit{lead} perturbation: cells are first exposed to hydrodynamic perturbation for 135 h (right after inoculation), thereafter transferred to static condition for rest of the experiment. Similarly, (F) 168 h lead refers to transfer from dynamic to static condition after 168 h.}
\label{Fig_schematic_rounds}
\end{figure}

In this study, we investigate how changes in fluid flow act as environmental cues that generate ecological memory in swimming phytoplankton. Phytoplankton are microscopic, photosynthetic organisms that underpin Earth’s biogeochemical cycles \cite{basu2018phytoplankton, morel2003biogeochemical, asselot2021missing}. Beyond primary production, phytoplankton interact with nitrogen-fixing cyanobacteria, viruses, and fungal symbionts within a nutrient-recycling microbial web \cite{jover2014elemental, khaliq2022arbuscular}. These interaction networks regulate nutrient regeneration, carbon remineralization, and the efficiency of the biological carbon pump, directly coupling phytoplankton community performance to climate-relevant carbon fluxes \cite{jiao2024microbial, zehr2011nitrogen, suttle2007marine}. Because of this tight coupling, understanding the ecological rules that govern phytoplankton performance is central to both scientific and societal needs \cite{behrenfeld2021phytoplankton}. 

By linking controlled hydrodynamic perturbations with quantitative measurements of growth kinetics, and phenotypic traits, we determine how mechanical history influences subsequent cellular and population-level responses. When gentle, hydrodynamic forcing, a abiotic factor in phytoplankton ecosystems, can enhance CO$_2$ and nutrient delivery, and reduce self-shading, whereas stronger flows can deform cells, disrupt flagellar propulsion, or truncate vertical migrations that are essential to motile taxa \cite{Sengupta2017Phytoplankton, Carrara2021Bistability, domozych2012cellwalls, kwok2023dinoflagellate, Sullivan2003}. At the scale of an individual motile cell---sub-micrometer to tens of micrometers---the aspect of hydrodynamic forcing that matters most is the local shear and rotation at Kolmogorov micro-scales \cite{sengupta2023planktonic}. These small-scale flows can align, reorient, or destabilise swimmers and alter the thickness of diffusive boundary layers \cite{Sengupta2017Phytoplankton, guasto2012fluid}. For gyrotactic and gravitactic taxa, background shear can bifurcate vertical migration, generate thin layers, or trap cells, demonstrating tight coupling between mechanical forcing, behaviour, and habitat use \cite{durham2009disruption}. Changes in stratification, wind, and sea-ice cover modulate the frequency and strength of such micro-scale shears, thereby linking climate-driven physical variability to the mechanical environment experienced by phytoplankton \cite{viljoen2024climate, muilwijk2024future, hopkins2021control}.


\begin{figure}[t]
\centering
\includegraphics[width=1\columnwidth]{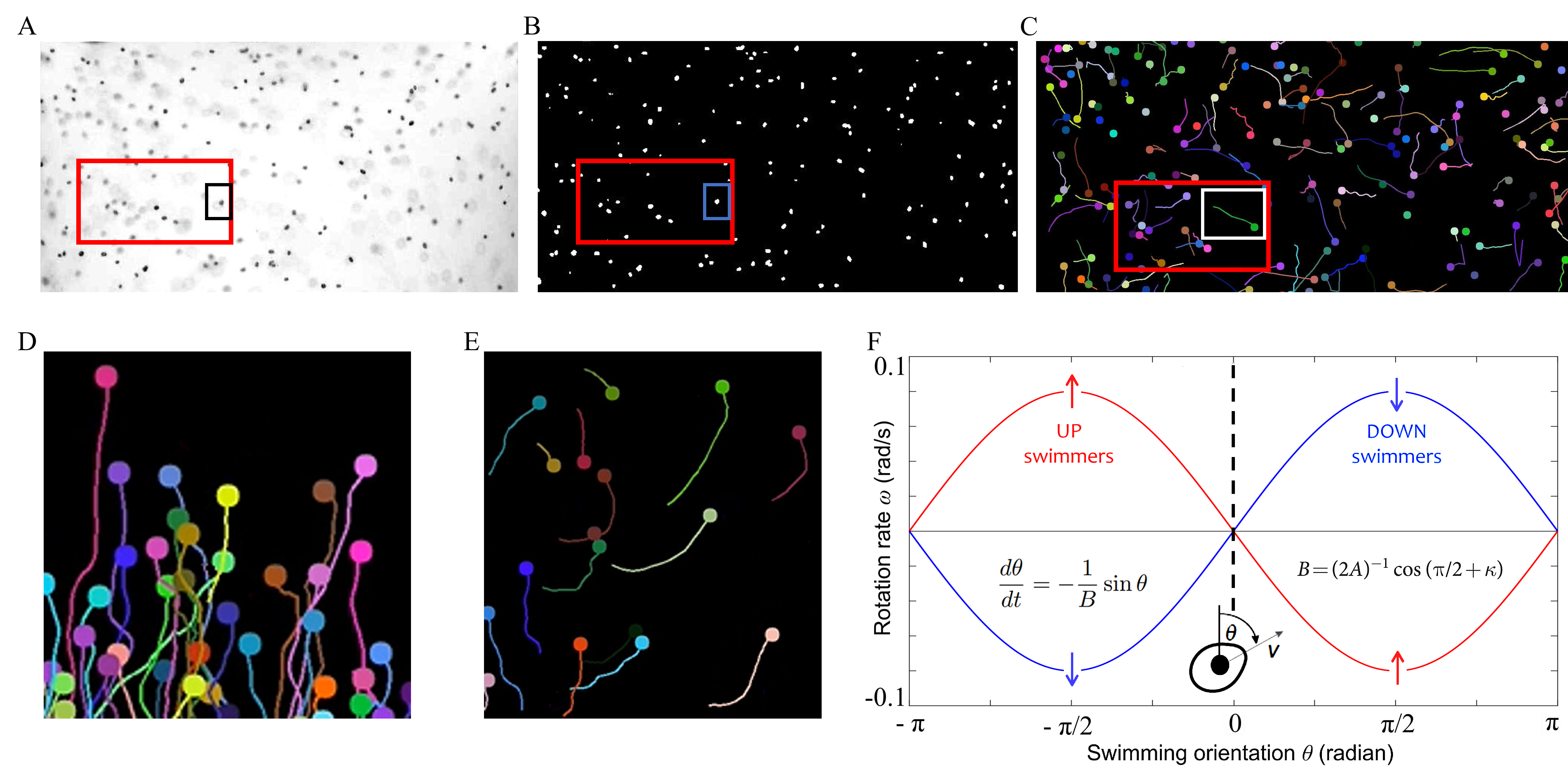}
\caption[Workflow for cell tracking and estimation of reorientation timescale of motile microorganisms.]{\textbf{Quantifying swimming and reorientation dynamics of motile microbes.} Raw (A) and binarized (B) images of swimming cells; insets present magnified view. (C) Swimming trajectories; inset presents zoom-in views of a selection of trajectories. 
(D) Trajectories visualized in the milifluidic chamber reveals diversity of trajectories and reorientation events (E), visualized as curved trajectories. (F) Instantaneous swimming angle ($\theta$), plotted against rotational velocity ($\omega$) provides reorientation time scales, following sinusoidal fitting. Representative reorientation plots of UP-swimming (negative gravitaxis) and DOWN-swimming cells (positive gravitaxis). Experimental data are fitted with a sinusoidal function ($\kappa$ is an imposed phase shift), which determines the reorientation timescale,($\tau_r$), from the best-fit sinusoid.}
\label{fig:supp1}
\end{figure} 


Phytoplankton can rapidly remodel gene expression, morphology, and motility \cite{collins2014evolutionary, anneville2018plasticity}, but plasticity has rate and magnitude limits beyond which acclimation becomes slow or costly \cite{botero2015evolutionary}. When external variability outpaces acclimation, populations may diversify phenotypes (bet-hedging) \cite{tang2024bet}, with epigenetic mechanisms stabilising alternative states across generations and accelerating community turnover \cite{reusch2013experimental, naselli2024analysis}. 
Responses to environmental change---including hydrodynamic cues---should therefore be interpreted not only in terms of forcing intensity, but also in relation to (i) the onset timing of forcing relative to physiological stage and (ii) the sequence in which the environmental cue is introduced (quiescent conditions followed by perturbations, or vice-versa), which together determine the extent of the functional response \cite{sengupta2023planktonic, stocker2012marine, main2014effects, powers2012sinking}. Within this framework, the motile raphidophyte \emph{Heterosigma akashiwo} is a tractable model for linking cell-scale mechanics to population-level outcomes \cite{Sengupta2017Phytoplankton, Carrara2021Bistability, durham2009disruption, durham2013turbulence, zeng2022sharp, Harvey2015, Anderson2022}. \emph{H.~akashiwo} combines flagellar motility with sensitive environmental sensing \cite{Sandoval-Sanhueza2022}, enabling diel vertical migrations tracking light and nutrient fields \cite{Sengupta2017Phytoplankton, Sengupta2022Active, thangaraj2023ocean}. Laboratory studies show that brief hydrodynamic pulses can reorient swimming paths, trigger intracellular signalling, and reorganise intracellular allocation \cite{kakavand2025hydrodynamic, Sengupta2017Phytoplankton, Cho2016Mechanotransduction}, while pH, temperature, salinity, and CO$_2$ further modulate these responses in non-linear ways \cite{Kim2013, qin2024examining, MehdizadehAllaf2024, Ghoshal2025}.

The timing of hydrodynamic forcing -- defined as the onset of the hydrodynamic cue relative to the growth stage of a cell population -- is a key factor in \textit{H. akashiwo} response \cite{kakavand2025hydrodynamic}. However, specific thresholds which distinguish beneficial hydrodynamic cues (those enhancing resource delivery, maintain swimming stability and growth) from the damaging ones (detrimental to behaviour and physiology) remain unclear. How these thresholds change across different growth stages is not yet understood. Furthermore, the role of prior exposure in modulating cellular sensitivity, including potential legacy effects, has not been fully elucidated. To address these, we map the stage-dependent tolerance, and explore how the temporal structure of forcing influence cellular responses. By focusing on hydrodynamic perturbations where local shear and vorticity directly affect motile cells and their near-field transport, we systematically examine the timing of a defined hydrodynamic perturbation across lag, exponential, and stationary phases of two distinct strains of \textit{H. akashiwo}, quantifying the resulting changes in growth and motility. Resolving these relationships help us to understand the plankton ecology in dynamic settings, thereby providing a framework for predicting microbial response to ecologically relevant cues. 


\section*{Results} 

\subsection*{Hydrodynamic history shapes growth kinetics of motile microorganisms}
\label{Impact of hydrodynamic perturbation timing on phytoplankton growth}

\subsubsection*{Standard scenario: Transition from static to hydrodynamic cues}
\label{Standard transitions from static to hydrodynamically perturbed conditions}
\begin{figure}[h]
\centering
\includegraphics[width=0.8\columnwidth]{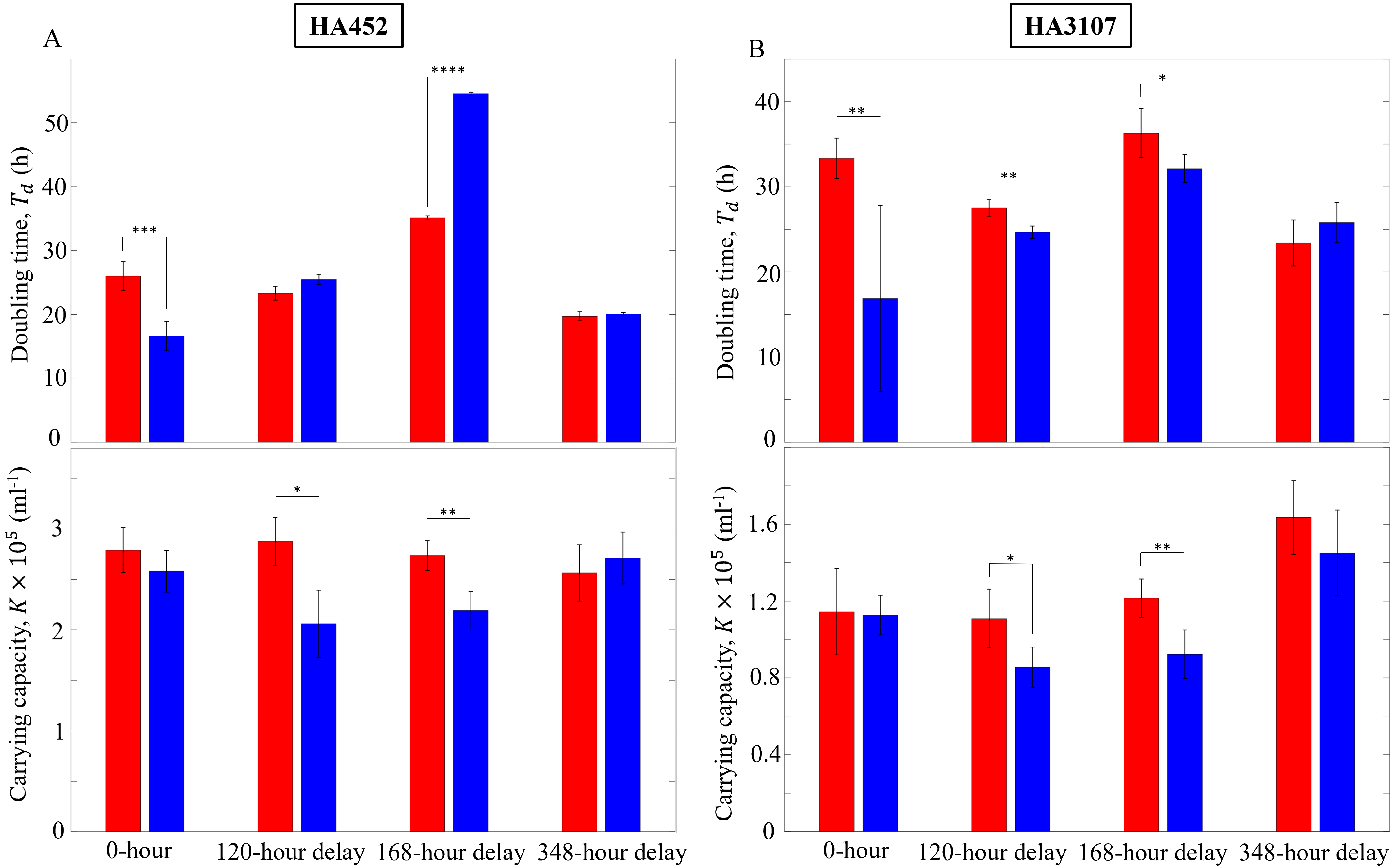}
\caption{{\textbf{Hydrodynamic perturbations impact doubling time and carrying capacity of motile microbes.} (A) Doubling time (\(T_d\), upper) and carrying capacity (\(K\), lower) from logistic fits for HA452 under forward hydrodynamic-perturbation scenarios, comparing static (red) and perturbed (blue) cultures. Early perturbation (0-hour) accelerates growth without altering \(K\), whereas perturbation imposed during exponential growth (120- and 168-h delay scenarios) reduces biomass yield and, at 168~h, prolongs \(T_d\); late perturbation (348~h) has no detectable effect. (B) As in (A) for HA3107, which shows generally stronger reductions in \(T_d\) and \(K\) at 120- and 168-h delays. Error bars denote standard deviations.  Statistical significance is indicated as follows: (*) for $P < 0.05$, (**) for $P < 0.01$, (***) for $P < 0.001$, and (****) for $P < 0.0001$.}}
\label{Fig_cc_dt_forward}
\end{figure}

Motile microorganisms including \textit{H.~akashiwo} -- our model microbe -- responds to hydrodynamic cues in a strain-specific and history-dependent manner \cite{Sengupta2017Phytoplankton, kakavand2025hydrodynamic}. The present study, conducted for a given hydrodynamic cue (imposed by the orbital shaker at 110 rpm), captures the effect of perturbation timing on the growth kinetics across different growth stages (Figure \ref{Fig_schematic_rounds}). Cell cultures were transferred from static to hydrodynamic cues at 0-hour, 120-hour, 168-hour, and 348-hour after their initial inoculation. The relative response was characterised using two complementary sets of metrics, namely impact on microbial swimming (Fig. \ref{fig:supp1}) and population growth kinetics (Fig. \ref{Fig_growth_452} for HA452 cells and Fig. \ref{Fig_growth_3107} for HA3107 cells), described by logistic fits (inset panel in each growth curve) that yielded the specific growth rate (doubling time) and carrying capacity (Fig. \ref{Fig_cc_dt_forward}).

For HA452, the 0-hour delay scenario, where hydrodynamic perturbation began immediately after inoculation, showed a substantial acceleration in population growth compared to the static control (Fig.~\ref{Fig_growth_452}A). The cultures reached stationary phase approximately 50~h earlier, with logistic modeling revealing a significantly shorter doubling time in the shaken cultures while carrying capacity remained comparable to that of the static controls (Fig.~\ref{Fig_cc_dt_forward}A and C). In the 120-hour delay scenario, when hydrodynamic perturbation was introduced after 120~h of static growth, the growth trajectories of perturbed cultures initially mirrored the static controls but later diverged. After approximately 130~h of shaking, perturbed cultures slowed and entered the stationary phase earlier, which was accompanied by a reduced carrying capacity (Fig.~\ref{Fig_cc_dt_forward}A and C). In the 168-hour delay scenario, the perturbation was applied after 168~h, at which point the cultures had entered mid-exponential growth. The immediate impact of hydrodynamic forcing was a considerable reduction in growth rate (Fig.~\ref{Fig_growth_452}C), with cultures transitioning from exponential to a more linear growth trajectory. Both doubling time and carrying capacity were significantly compromised in the perturbed cultures (Fig.~\ref{Fig_cc_dt_forward}A and C). 

\begin{figure}[H]
    \centering
    \includegraphics[width=0.8\textwidth]{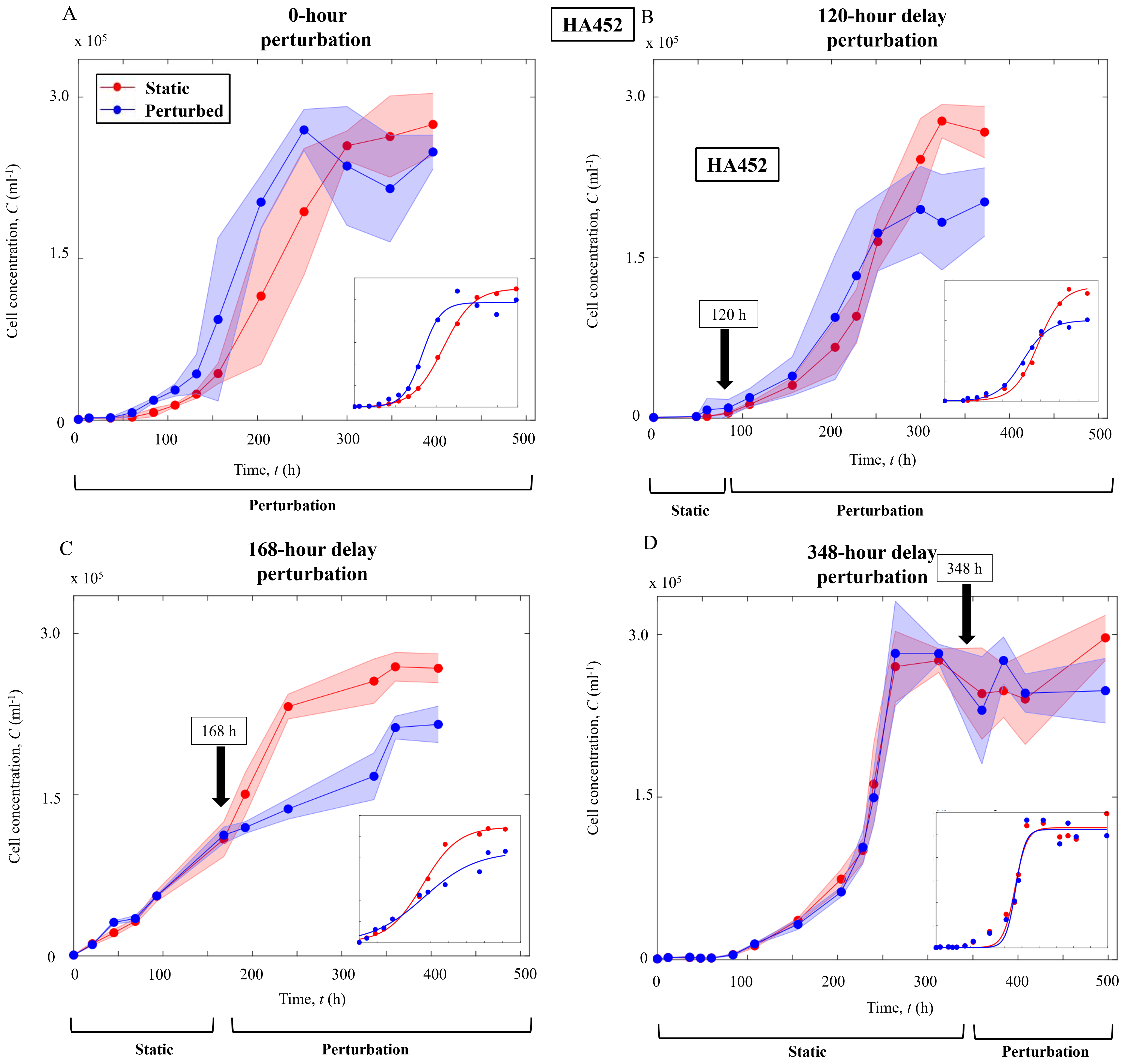}
\end{figure}
\captionof{figure}[Impact of hydrodynamic perturbation timing on \textit{H. akashiwo} (CCMP452) growth]{\textbf{Impact of standard hydrodynamic scenarios (static-to-perturbed) on the growth of HA452 population.} (A)--(D): Growth curves of HA452 cells for both the experimental perturbed samples and the static control samples under the following perturbation scenarios: (A) 0-hour, (B) 120-hour delay, (C) 168-hour delay, and (D) 348-hour delay. In all panels, the inset displays the logistic models fitted to the corresponding growth curves. For each data point, two biological replicates and two technical replicates were measured. In the growth curves, the standard deviation is represented by the shaded region surrounding the solid curve, which shows the mean values.}
\label{Fig_growth_452}

For HA3107, the growth responses closely mirrored those observed for HA452, albeit with stronger negative effects at later onsets. In the 0-hour delay scenario, where perturbation began immediately after inoculation, the growth enhancement had a comparable effect (Fig.~\ref{Fig_growth_3107}A). Logistic analysis revealed that HA3107 exhibited faster growth and a comparable carrying capacity to the static control, for at least 9 days. In the 120-hour delay scenario, HA3107 exhibited more pronounced impairment than HA452, with both reduced growth rate and carrying capacity in the perturbed cultures compared to static controls (Fig.~\ref{Fig_growth_3107}B). In the 168-hour delay scenario, HA3107 exhibited similar but more severe growth reduction compared to HA452, with pronounced decline in both specific growth rate and carrying capacity after perturbation (Fig.~\ref{Fig_growth_3107}C). This supports the notion that HA3107 is more vulnerable to perturbations applied during mid-exponential growth.

\begin{figure}[H]
    \centering
    \includegraphics[width=0.85\textwidth]{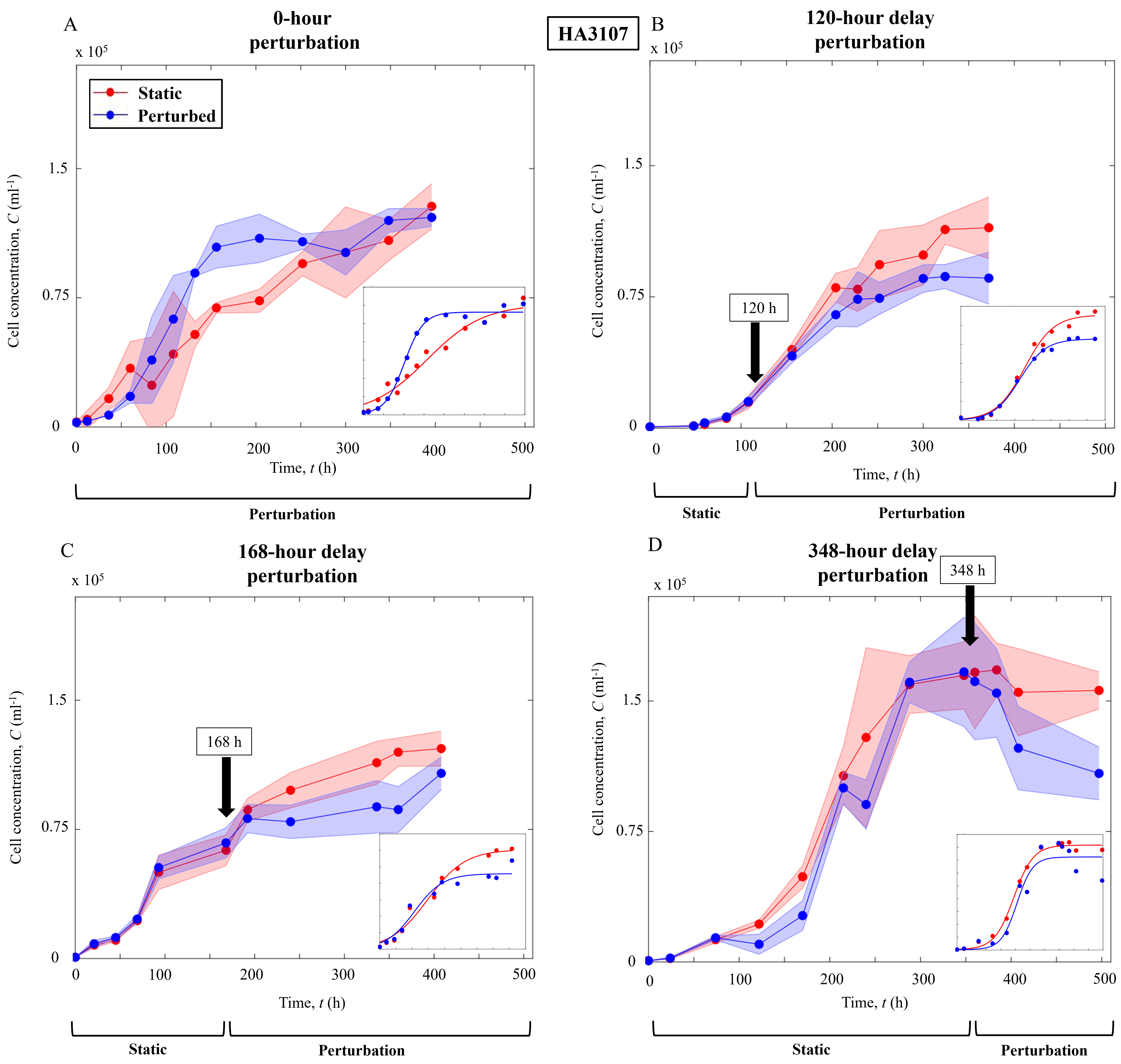}
\end{figure}
\captionof{figure}[Impact of hydrodynamic perturbation timing on \textit{H. akashiwo} (CCMP3107) growth]{\textbf{Impact of standard hydrodynamic scenarios (static-to-perturbed) on the growth of HA3107 population.} (A)--(D): Growth curves of HA3107 cells for both experimental perturbed samples and static control samples under the following perturbation scenarios: (A) 0-hour, (B) 120-hour delay, (C) 168-hour delay, and (E) 348-hour delay. In all panels, the inset displays the logistic models fitted to the corresponding growth curves.  Each data point represents two biological replicates and two technical replicates. For the growth curves, the shaded regions around the solid curves represent the standard deviations.}
\label{Fig_growth_3107}

Finally, in the 348-hour delay scenario, where perturbation was applied after cultures reached stationary phase, both HA452 and HA3107 exhibited no significant change in logistic growth parameters, as expected (Figs.~\ref{Fig_growth_3107}D and \ref{Fig_cc_dt_forward}B). However, both strains entered the death phase earlier under hydrodynamic perturbation than their static controls, indicating that even stationary-phase cells are affected by external mechanical forcing, albeit without impacting their core growth metrics. These findings confirm that the timing of hydrodynamic perturbation during the growth cycle is a critical determinant of fitness in \textit{H.~akashiwo}. While early and continuous exposure to hydrodynamic forcing supports high fitness in both strains, delaying perturbation until later growth stages causes a progressive loss of biomass. The results also highlight strain-specific differences in response to perturbation, with HA3107 showing a more pronounced vulnerability to early and mid-exponential exposure, and a shorter survival window under late-onset perturbation.

\FloatBarrier

\subsubsection*{Reverse transitions from hydrodynamically perturbed to static conditions}
\label{Backward transitions from hydrodynamically perturbed to static conditions}

 Reverse scenarios, i.e., perturbed-to-static conditions, where cultures were first exposed to hydrodynamic perturbation and then returned to static conditions, allowed us to investigate how prior exposure to hydrodynamic cues affects microbial behaviour and physiology once the cue ceases to act. Specifically, we considered two reverse scenarios in which cultures were kept under orbital shaking from inoculation and then shifted back to static conditions at defined times corresponding to early and mid–exponential growth (135-hour and 168-hour reverse scenarios). Together with the continuously shaken 0-hour delay treatment, these experiments allow us to assess how cells conditioned under hydrodynamic perturbation perform when transitioning back to static conditions and how this transition influences growth dynamics and swimming performance in both HA452 and HA3107.

The 135-hour and 168-hour reverse scenarios were designed so that population densities at the time of the transition closely matched those in the standard 120-hour and 168-hour delay scenarios, respectively, as inferred from the growth curves. This design facilitated direct comparison between standard and reverse scenarios at comparable physiological states. For HA452, the 135-hour reverse scenario involved continuous shaking from inoculation until early exponential phase, followed by a return to static conditions at 135~h. Up to the transition point, perturbed and static cultures showed very similar growth trajectories (Fig.~\ref{Fig_growth_reverse_all}A). Immediately after the shift to static conditions, the previously shaken cultures displayed a brief secondary lag phase, visible as a temporary reduction in growth slope relative to the always-static controls, before recovering to an exponential regime. Logistic fits revealed that the doubling time in the reverse-treated cultures was not significantly different from that of the static controls, but the carrying capacity was modestly reduced. This suggests that early hydrodynamic exposure followed by a return to static conditions does not permanently alter the growth kinetics but slightly lowers the maximum biomass that can be sustained. 


In the 168-hour reverse scenario, cultures were shaken from inoculation until mid–exponential phase and then returned to static conditions. The last measurement at 168~h under continuous perturbation showed that growth (Fig.~\ref{Fig_growth_reverse_all}B) was still comparable between treatments at the moment of transition. However, after the shift to static conditions, the trajectories diverged. The reverse-treated cultures displayed a pronounced secondary lag and a slower approach to stationary phase compared with the always-static controls. Logistic fits indicated that the reverse-treated cultures had a shorter fitted doubling time but a significantly reduced carrying capacity. In other words, cells which had experienced prolonged early perturbation divided more rapidly during the recovered exponential segment, but the population saturated at a lower density. 

\begin{figure}[H]
    \centering
    \includegraphics[width=0.8\textwidth]{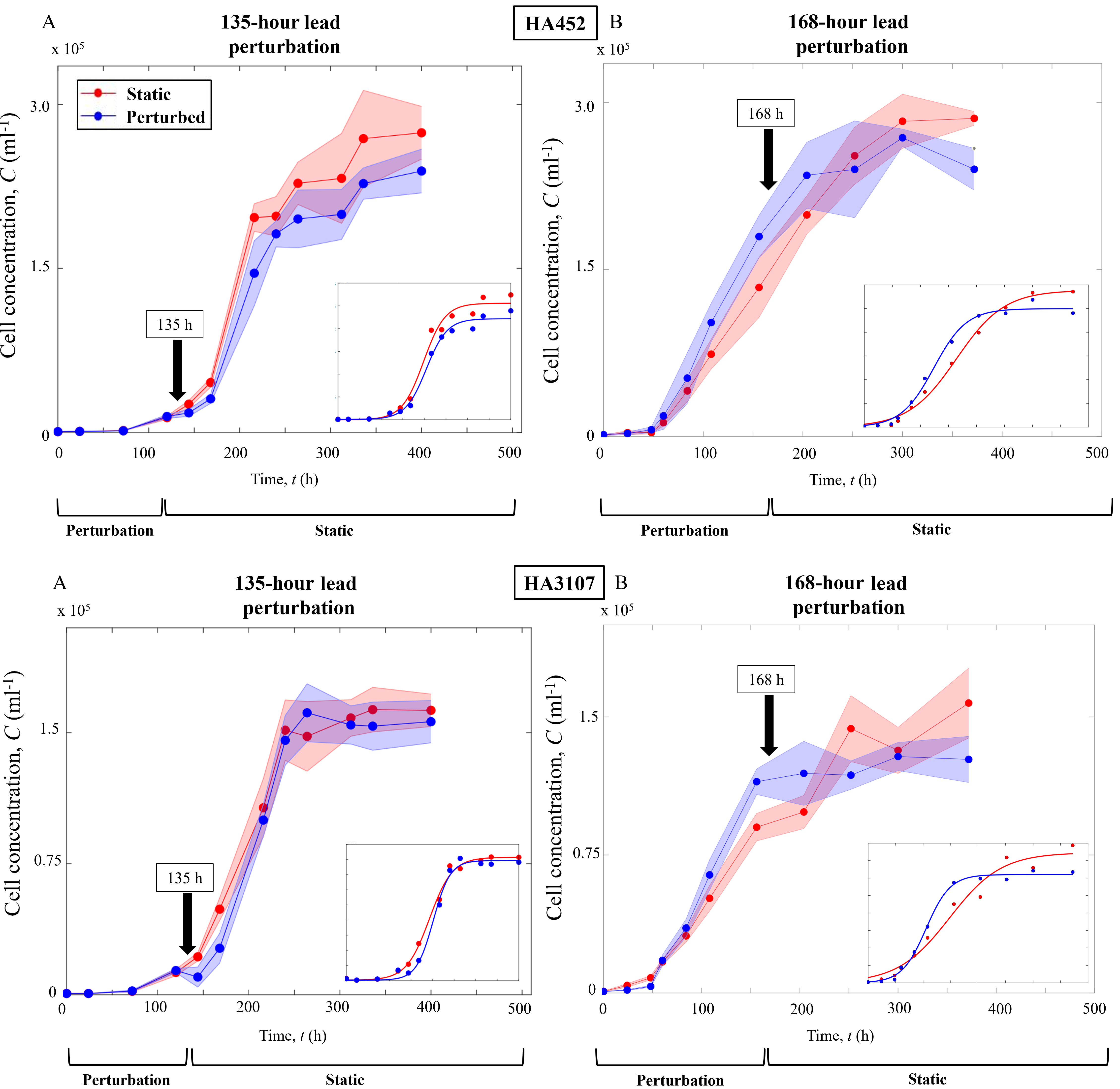}
\end{figure}
\captionof{figure}[Impact of reverse hydrodynamic perturbation on \textit{H. akashiwo} (CCMP452) growth]{\textbf{Impact of reverse hydrodynamic perturbation (perturbed-to-static) scenarios on growth of motile microbes.} (A)–(B) Growth curves of HA452 cells for the 135-hour lead and 168-hour lead reverse scenarios, respectively. (C)–(D) Growth curves of HA3107 cells for the same reverse scenarios, with static controls (red) and reversely perturbed cultures (blue). Insets in all panels show logistic fits used to extract doubling time and carrying capacity. Each data point represents two biological replicates and two technical replicates; shaded regions around the mean curves indicate standard deviations.}
\label{Fig_growth_reverse_all}

Taken together, the reverse scenarios in HA452 show that early hydrodynamic exposure continues to influence fitness even after perturbation is removed. When the transition back to static conditions occurs early, populations retain a near-normal division rate but pay a moderate cost in carrying capacity and rely more on non-photochemical dissipation (data not reported here). When the same transition is delayed to mid-exponential phase, the effect of prior mechanical exposure becomes more pronounced, leading to a stronger reduction in biomass yield and a deterioration of photopjhysiology. These results, in combination with the standard scenarios, indicate that the direction and timing of transitions between static and perturbed environments significantly affect the fitness outcomes of the organism.

For HA3107, two distinct reverse transitions from hydrodynamically perturbed to static conditions were designed: the 135-hour and 168-hour reverse scenarios, as in the case of HA452. These times were chosen to match the cell densities at the moment of release from perturbation with those at which hydrodynamic perturbation was imposed in the standard 120-hour and 168-hour delay scenarios, respectively. This design allows backward and forward experiments to be compared at matched population states.

\begin{figure}[H]
    \centering
    \includegraphics[width=0.8\columnwidth]{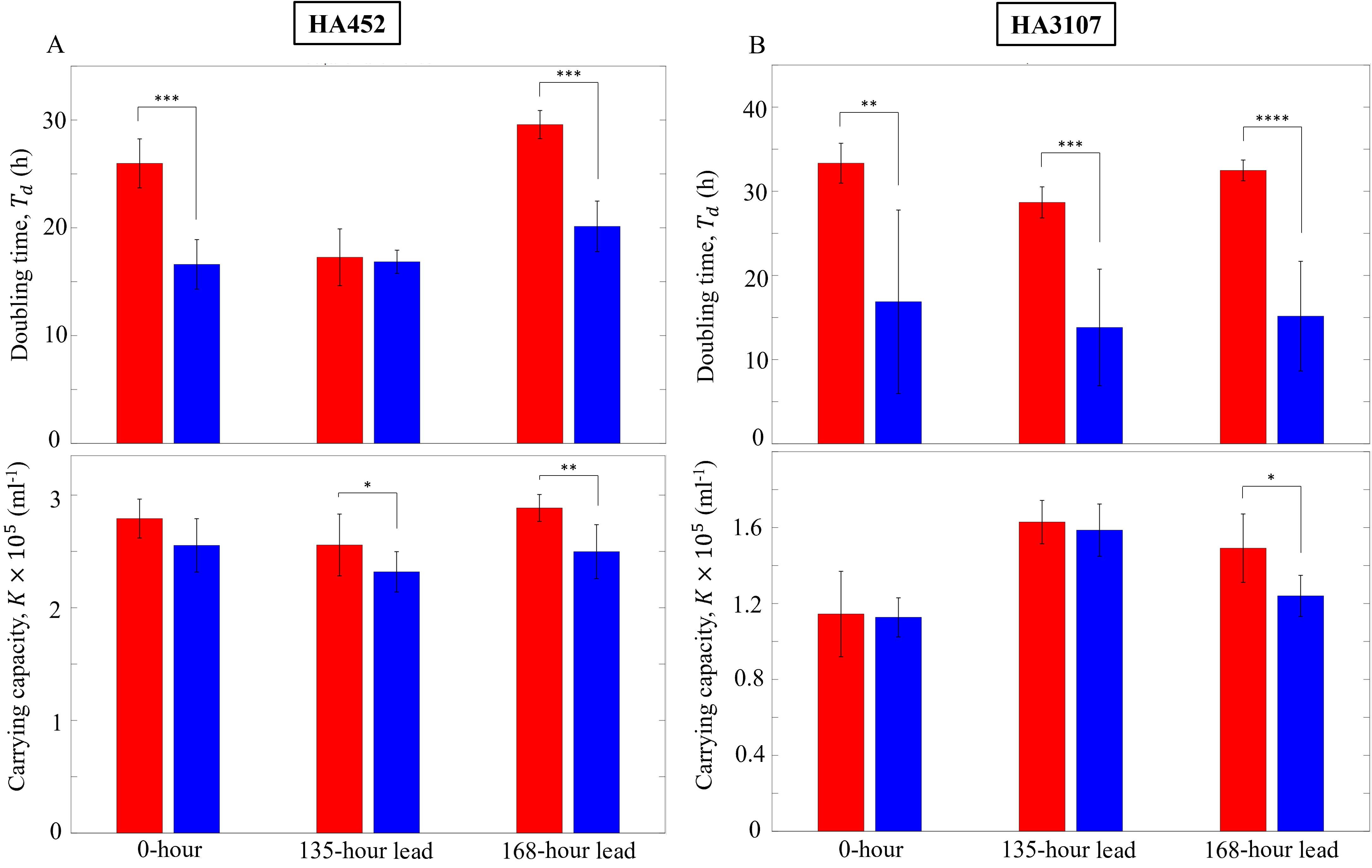}
    \caption[Reverse hydrodynamic history effects on doubling time and carrying capacity in \textit{H. akashiwo}]{\textbf{Reverse hydrodynamic perturbation impacts doubling time and carrying capacity.} (A) Doubling time (\(T_d\), upper) and carrying capacity (\(K\), lower) from logistic fits for HA452 under the 0-hour and reverse scenarios (135-hour lead and 168-hour lead), comparing static (red) and reverse-perturbed (blue) cultures. (B) As in (A), data for HA3107. Error bars denote standard deviations; asterisks indicate significant differences between static and perturbed cultures (two-sided \(t\)-test).}
    \label{Fig_cc_dt_reverse}
\end{figure}

In the 135-hour reverse scenario, growth trajectories of HA3107 initially followed those of continuously shaken cultures, reflecting the shared early history under orbital agitation. After the shift to static conditions, the perturbed cultures displayed a mild secondary lag and then resumed growth, but with a slower increase in cell concentration than the static controls (Fig.~\ref{Fig_growth_reverse_all}C). Logistic fits revealed that the doubling time of the reverse-history cultures was significantly longer than that of the purely static cultures, while the carrying capacity remained similar (Fig.~\ref{Fig_cc_dt_reverse}B). A finite period of early hydrodynamic perturbation followed by a return to static conditions thus slowed population growth but did not substantially alter final biomass yield in HA3107. In the 168-hour reverse scenario, a stronger imprint of hydrodynamic history was observed. Up to the transition time at 168~h, reverse-history and continuously shaken cultures showed similar trajectories. However, after the switch to static conditions, the growth of reverse-history cultures decelerated strongly relative to static controls and reached a lower plateau (Fig.~\ref{Fig_growth_reverse_all}D). Logistic analysis showed that both doubling time and carrying capacity were significantly reduced in the reverse-history cultures compared with purely static cultures (Fig.~\ref{Fig_cc_dt_reverse}B). A finite period of agitation that extends deeper into exponential growth thus leads to a lasting reduction in both the rate and extent of biomass accumulation.

This pattern indicates that, once the mechanical environment is reversed late in exponential growth, HA3107 finds it difficult to maintain fitness and compensates by diverting a larger fraction of absorbed energy into regulated dissipation, in line with the reduced growth rate and lower carrying capacity. Comparison of the 135-hour and 168-hour reverse scenarios thus shows a clear timing dependence in the reverse direction. Early release from perturbation leaves HA3107 with modestly slower growth but preserves carrying capacity, whereas a later release produces larger reductions in both growth rate and biomass yield. In combination with the standard scenarios, these results indicate that HA3107 retains a strong effect of its hydrodynamic history and that the fitness penalty associated with a transition from perturbed to static conditions becomes more severe as the transition is delayed further into the exponential phase.

\subsection*{Hydrodynamic perturbations reprogram microbial motility, when applied prior to exponential growth phase}
\label{motility_results}

Phytoplankton fitness in fluctuating aquatic environments depends strongly on the ability of cells to adjust motility when light, nutrients, or fluid motion change. Previous work showed that \textit{Heterosigma akashiwo} rapidly reshapes its motility under nutrient stress by altering cell size, morphology, intracellular mass distribution, and flagellar beating patterns \cite{Sengupta2022Active}. As a strongly gravitactic species, \textit{H.\ akashiwo} uses this phenotypic plasticity to regulate its vertical position by combining gravitational sensing with external cues such as light and chemical gradients. Hydrodynamic conditions therefore act not only as a mechanical constraint but also as a control parameter that can reshape swimming behaviour and, ultimately, ecological performance \cite{sengupta2023planktonic}.

To place the experimental perturbation into a turbulence framework, the energy dissipation rate in the orbital shaker was estimated as $\epsilon \approx 9.54 \times 10^{-4}\,\mathrm{m}^2\,\mathrm{s}^{-3}$ (see Methods), roughly one order of magnitude higher than typical dissipation levels experienced by phytoplankton in open-ocean turbulence \cite{Sengupta2017Phytoplankton}. The corresponding Kolmogorov length scale, \(\big(\eta = (\nu^3/\epsilon)^{1/4}\big)\), is about $0.18~\mathrm{mm}$, and the associated shear rate, \(\big(S_\eta = (\epsilon/\nu)^{1/2}\big)\), is $\sim 30~\mathrm{s}^{-1}$, giving a characteristic fluctuation time scale, \(\big(\tau_\eta = 1/S_\eta\big)\), of about $0.03~\mathrm{s}$. Cell diameters are $\sim 10\text{--}20~\mu\mathrm{m}$, much smaller than the Kolmogorov scale, so each cell experiences the imposed flow as an approximately linear velocity gradient across its body. This regime is well suited for probing how motility traits respond to controlled small-scale shear.

\begin{figure}
\centering
\includegraphics[width=1\columnwidth]{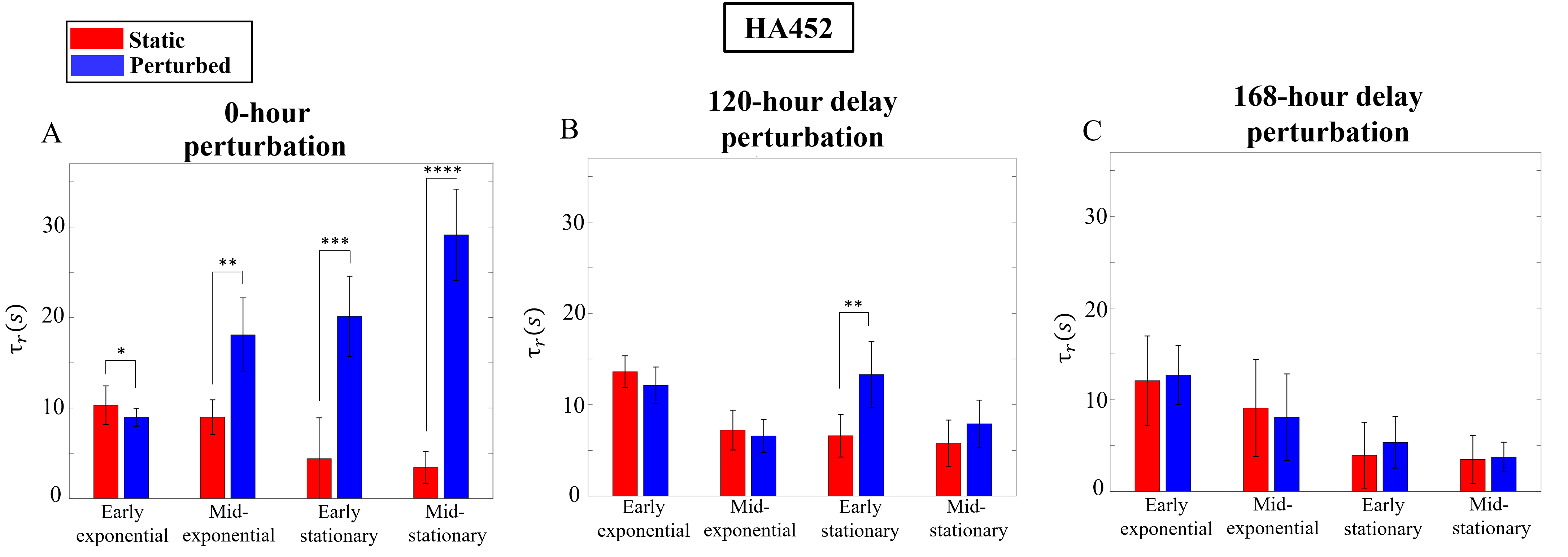}
\caption[Adjustment of reorientation timescales of HA452 due to hydrodynamic cues]{{\textbf{Adjustment of reorientation timescales of HA452 due to hydrodynamic cues (standard perturbation scenarios).} {(A) Reorientation timescales of HA452 cells under the 0-hour perturbation scenario, at early exponential, mid-exponential, early stationary, and mid-stationary growth phases. Early and continuous hydrodynamic perturbation significantly impaired the reorientation time (and thus, the swimming stability) of the cells. (B) Reorientation timescales of cells for the 120-hour delay scenario. Hydrodynamic perturbation introduced after the cells had reached the exponential phase did not affect their swimming stability. In case of any impact, the cells were able to recover their stability. (C) Reorientation timescales of cells under the 168-hour delay scenario. Delayed perturbation did not impair the swimming stability of HA452 cells. In all plots, the mean reorientation timescale of two biological and two technical replicates is presented, with standard deviation represented as error bars. Statistical significance is denoted as follows: (*) for $P < 0.05$, (**) for $P < 0.01$, (***) for $P < 0.001$, and (****) for $P < 0.0001$.}}}
\label{Fig_time_orient_comparison_452}
\end{figure}

\begin{figure}[h]
\centering
\includegraphics[width=0.9\columnwidth]{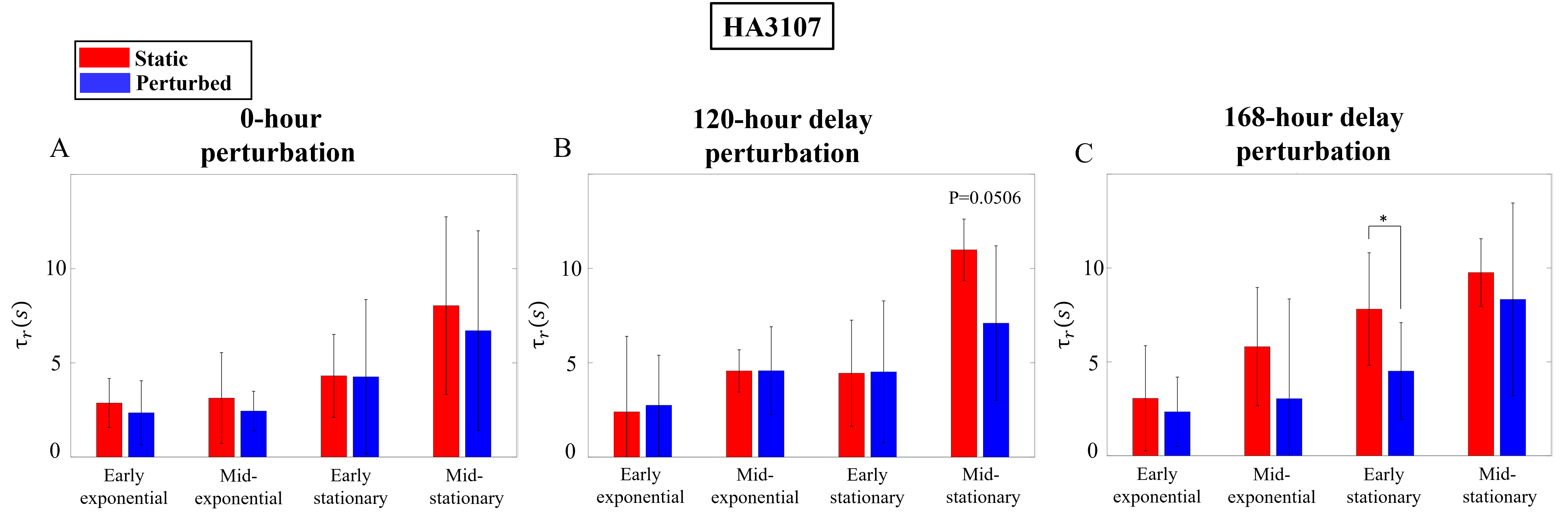}
\caption[Comparison of reorientation timescales between static and perturbed \textit{H. akashiwo} (CCMP452) cells across various perturbation scenarios]{{\textbf{Comparison of reorientation timescales under standard perturbation scenarios
for HA3107. } {(A) Reorientation timescales of HA3107 cells under the 0-hour delay perturbation scenario at four distinct time points: early exponential, mid-exponential, early stationary, and mid-stationary growth phases. Early and continuous hydrodynamic perturbation did not significantly affect the swimming stability of the cells. (B) Reorientation timescales of cells under the 120-hour delay scenario. Hydrodynamic perturbation applied after the cells had reached the exponential phase showed no effect on their swimming stability. (C) Reorientation timescales of cells under a 168-hour delay scenario. In all plots, the mean reorientation timescale of two biological and two technical replicates is presented, with standard deviation shown as error bars. Statistical significance is indicated as follows: (*) for $P < 0.05$, (**) for $P < 0.01$, (***) for $P < 0.001$, and (****) for $P < 0.0001$.}}}
\label{Fig_reorient_comparison_3107}
\end{figure}

Motility was quantified with three complementary descriptors (Figure \ref{fig:supp1}). First, the mean vertical swimming velocity, \(v_y\), measured across four growth phases (early exponential (48-72 h post-inoculation), mid exponential (120-144 h post-inoculation), early stationary (200-248 h post-inoculation), and mid stationary (350-398 h post-inoculation)) for static and perturbed cultures, with upward motion taken as positive (Figs. \ref{vertical_452_forward}, \ref{vertical_452_backward}). Second, the full distribution of swimming directions was resolved by constructing wind-rose plots that map the angular statistics of single-cell displacement vectors onto the unit circle (Figs. \ref{Fig_windrose_452} and \ref{Fig_windrose_3107}). For each trajectory, the instantaneous swimming direction was defined as the unit vector \(\hat{\boldsymbol{p}}(t) = \Delta \boldsymbol{r}(t)/\lvert \Delta \boldsymbol{r}(t) \rvert\), where \(\Delta \boldsymbol{r}(t)\) is the displacement over a fixed time interval. Third, swimming stability was quantified through a characteristic reorientation time, \(\tau_c\), for a gravitactic population \cite{Sengupta2017Phytoplankton}. For each condition, we tracked individual cell trajectories during relaxation after a chamber flip, computed the instantaneous swimming orientation \(\theta(t)\) with respect to the vertical and the associated angular velocity \(\omega(t) = \mathrm{d}\theta/\mathrm{d}t\), and then constructed the population-averaged curve \(\omega(\theta)\). This drift was fitted with a sinusoidal form \(\omega(\theta) \simeq A \cos(\theta + \kappa)\), as expected from the balance of gravitational and viscous torques in a gravitactic regime. Linearizing this relation around the stable orientation \(\theta_\mathrm{s}\) gives \(\mathrm{d}\theta/\mathrm{d}t \approx -(\theta - \theta_\mathrm{s})/\tau_r\), so that the local slope of \(\omega(\theta)\) near \(\theta_\mathrm{s}\) is \(-1/\tau_r\). From the fitted sinusoid, this slope is \(2A\), and the characteristic reorientation time is therefore obtained as \(\tau_r = 1/(2A)\) (Fig. \ref{fig:supp1}) \cite{Sengupta2017Phytoplankton}. Characteristic reorientation time \(\tau_r\) is therefore the time a gravitactic cell needs to realign its swimming direction back to its stable orientation after that orientation has been disturbed. In our analysis and plots we report \(\tau_r\) (Figs. \ref{Fig_time_orient_comparison_452} and \ref{Fig_reorient_comparison_3107}) as: smaller \(\tau_r\) corresponds to stronger orientational stability (faster realignment toward the stable swimming direction), whereas larger \(\tau_r\) indicates weaker stability.

In the 0-hour delay scenario, motility of HA452 population was persistently altered. Static controls increased their upward swimming velocity as growth progressed, reaching mean vertical speeds of order $80\text{--}100~\mu\mathrm{m}\,\mathrm{s}^{-1}$ by mid stationary phase, consistent with increasingly ballistic upward migration. Perturbed cultures showed strongly reduced vertical velocities at all stages (Fig.~\ref{vertical_452_forward}A), often close to zero during early exponential growth and never approaching the control values even at later phases.

Directional statistics mirror this suppression. Wind-rose plots for the 0-hour delay scenario (left column in Fig.~\ref{Fig_windrose_452}) show that static cells retained a narrow upward lobe concentrated around the vertical axis, whereas perturbed cells displayed almost isotropic swimming already in early exponential phase. The loss of a preferred upward direction was not compensated by enhanced horizontal exploration but rather accompanied by an overall reduction in swimming behaviour.

The reorientation-time analysis captures this loss of directional stability in a single metric. In the 0-hour delay scenario, the characteristic reorientation time of perturbed cells increased with growth and reached values several times larger than those of static controls (Fig.~\ref{Fig_time_orient_comparison_452}A). Static cultures, in contrast, exhibited decreasing reorientation times as they progressed from exponential to stationary phase, indicating more stable gravitactic alignment. Early and continuous hydrodynamic forcing therefore drives HA452 cells from a regime of directed upward motility with short reorientation times toward a regime of slow, nearly diffusive swimming with long-lived orientation fluctuations.

When hydrodynamic perturbation was introduced after an initial period of static growth, motility remained more robust. In the 120-hour delay scenario, vertical velocities of control and experimental cultures were comparable before the onset of agitation, as designed. After the switch to orbital shaking in early exponential phase, vertical velocities of perturbed cells initially decreased relative to static controls but progressively recovered so that by mid stationary phase the two conditions showed comparable mean values (Fig.~\ref{vertical_452_forward}B).

Analysis of swimming direction distribution for the 120-hour delay scenario (middle column in Fig.~\ref{Fig_windrose_452}) reveal a temporary broadening of the upward lobe in perturbed cultures, consistent with a transient increase in directional variability. However, the strong anisotropic upward bias characteristic of static controls was largely re-established at later times. Reorientation times showed a similar pattern. A modest increase in the characteristic time scale was detected shortly after the onset of hydrodynamic forcing, indicating a short-lived reduction in orientational stability, but values subsequently converged toward those of static cultures (Fig.~\ref{Fig_time_orient_comparison_452}B). These measurements show that \textit{H.\ akashiwo} can partly buffer or reverse motility disruptions when hydrodynamic forcing begins after cells have already entered exponential growth.

\begin{figure}[H]
    \centering
    \includegraphics[width=1\textwidth]{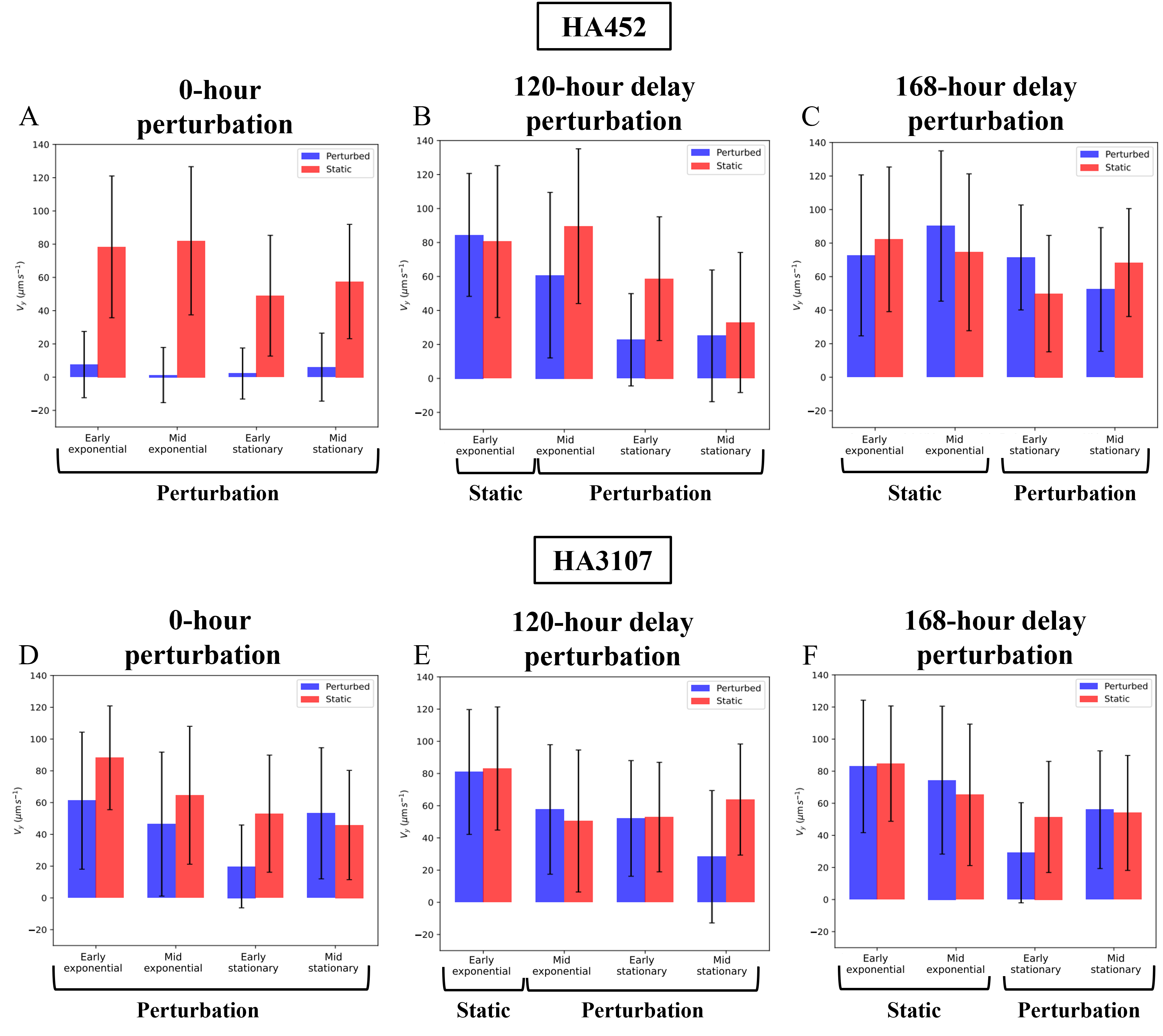}
\end{figure}
\captionof{figure}{\textbf{Ecological memory of hydrodynamic cues on motility traits under standard scenarios.} Vertical speeds of perturbed (blue) and static controls (red) cell cultures across different growth phases (early exponential, mid-exponential, early stationary, and mid-stationary). Panels (A–C) correspond to HA452 under the 0-hour (A), 120-hour (B), and 168-hour (C) delay scenarios, while panels (D–F) show the corresponding conditions for HA3107 (0-hour: D; 120-hour: E; 168-hour: F). In HA452 cells, immediate perturbation (0-hour) markedly reduced upward swimming relative to controls, whereas delayed perturbations produced minimal overall changes, with partial recovery and instances of higher velocities at certain phases. In contrast to strain HA452, perturbation timing does not alter the vertical swimming of HA3107 cells, with velocities remaining comparable to static controls. Values indicate mean speeds from two independent biological replicates and two technical replicates per time point.}
\label{vertical_452_forward}
\FloatBarrier

Hydrodynamic perturbation introduced at $168$ h post-inoculation, when cultures had already progressed deep into exponential growth and approached stationary conditions, produced only minor changes in HA452 motility. Vertical velocities of perturbed and static cells remained very similar across all subsequent time points (Fig.~\ref{vertical_452_forward}C). Directional distributions preserved a pronounced upward bias in both conditions, with only subtle broadening of the angular spread in the perturbed cultures (right column in Fig.~\ref{Fig_windrose_452}). Reorientation times also overlapped within experimental uncertainty (Fig.~\ref{Fig_time_orient_comparison_452}C), indicating that the stability of gravitactic alignment was largely insensitive to late-stage hydrodynamic forcing. Across the standard scenarios, the timing of hydrodynamic forcing along the growth curve emerges as a key control parameter for motility plasticity in motile species. Forcing from inoculation produces a long-lived reprogramming of motility, with strong suppression of upward swimming, loss of anisotropic orientation, and greatly increased reorientation times. Forcing initiated in early exponential phase causes only transient motility disruption followed by recovery, whereas hydrodynamic perturbation initiated in late exponential or early stationary phase barely alters motility. These observations indicate that early hydrodynamic history leaves a lasting imprint on HA452 motility, while cells exposed to forcing only after growth has been established respond in a far more reversible manner.

Under the same set of forward perturbations, strain HA3107 showed a contrasting motility phenotype compared with HA452. In the 0-hour delay scenario HA3107 cells maintained a relatively robust upward bias in their motility. Vertical swimming velocities of perturbed cultures decreased during early exponential growth compared with static controls, but the reduction was moderate and mean velocities converged toward control values by early and mid stationary phases. Wind-rose plots for this scenario show that static cultures developed a narrow upward lobe similar to HA452, whereas perturbed HA3107 cells preserved a clearly anisotropic distribution with a dominant upward sector, albeit with somewhat broader angular spread (left column in Fig.~\ref{Fig_windrose_3107}). Reorientation times differed only slightly between conditions and in some cases were marginally shorter in perturbed cultures at later stages (Fig.~\ref{Fig_reorient_comparison_3107}A), which points to preserved or even slightly enhanced orientational stability despite continuous exposure to shear. These measurements indicate that, unlike HA452, HA3107 can accommodate early hydrodynamic forcing without collapsing its gravitactic migratory state.

\newpage

\begin{figure}
    \centering
    \includegraphics[width=0.7\textwidth]{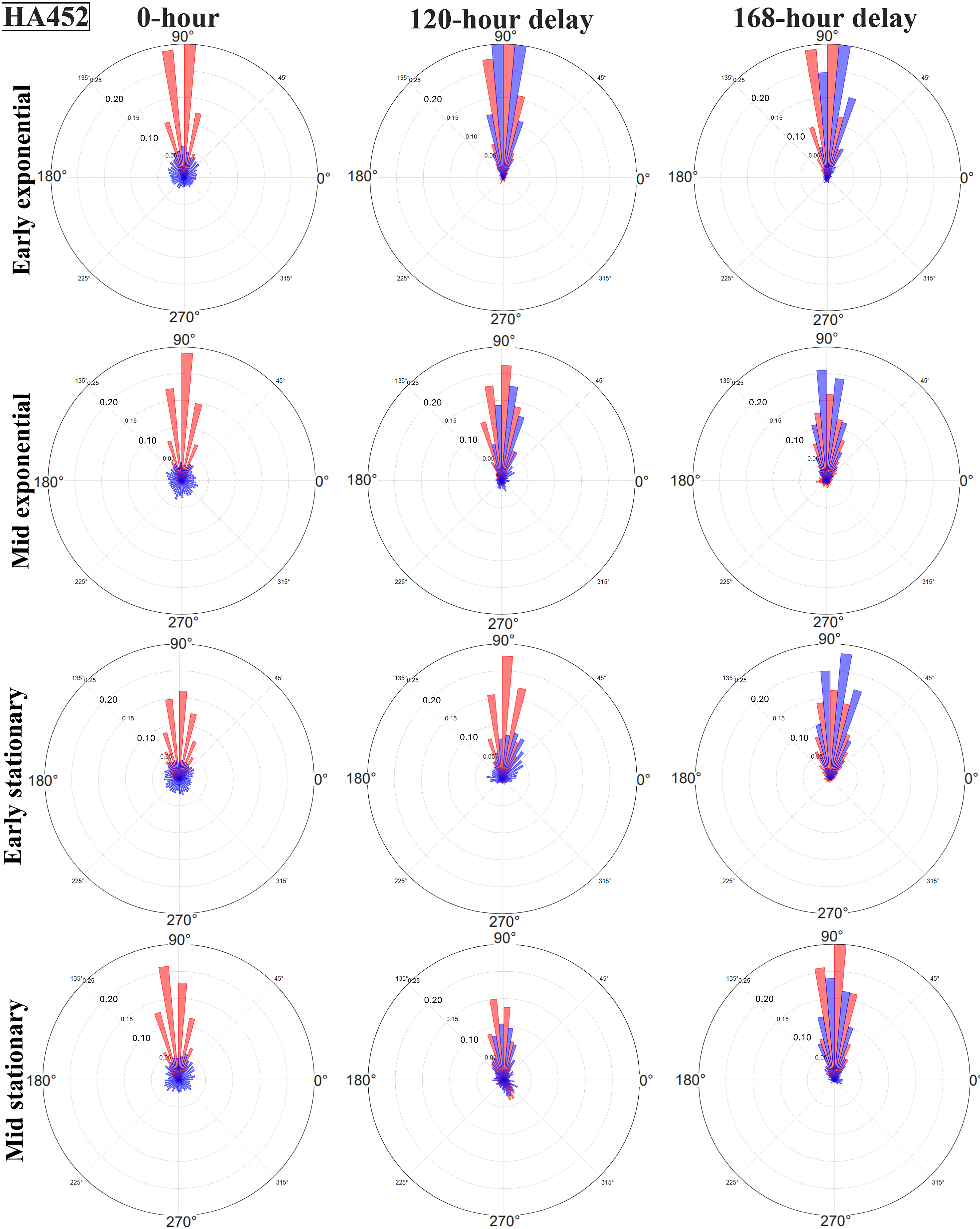}
\end{figure}

    \caption[Distribution of swimming directions of \textit{H. akashiwo} (CCMP452) cells under forward perturbation scenarios]%
    {{\textbf{Distribution of swimming directions of HA452 (standard scenarios).} 
    Windrose plots depict the swimming direction of HA452 cells for static cells (red) and perturbed cells (blue) under the 0-hour delay, 120-hour delay, and 168-hour delay scenarios. 
    The length of each bar is proportional to the relative frequency of cells swimming in that direction, with the relative frequencies indicated by the labels on the concentric circles. 
    Each row represents a specific time point across all perturbation scenarios, while each column corresponds to one of the three forward perturbation scenarios at the indicated times. 
    The swimming direction of HA452 cells is strongly modified under the 0-hour scenario, whereas delayed hydrodynamic perturbation (120- and 168-hour delay) does not alter the upward anisotropic swimming pattern. 
    Data for each plot is obtained from two biological and two technical replicates.}}
    
    \label{Fig_windrose_452}

When hydrodynamic perturbation was initiated at 120 hours, once HA3107 cultures had already entered exponential growth, motility impairments were again transient. Prior to agitation, vertical velocities of control and experimental cultures were comparable by design. After the onset of shaking, mean upward speeds of perturbed cells declined with a delay of several days and then recovered during mid stationary phase so that final velocities were comparable between treatments. Directional statistics mirror this delayed and reversible response. Swimming direction distribution plots for the 120-hour delay scenario reveal a temporary broadening and weakening of the upward lobe in perturbed cultures during early stationary phase, followed by restoration of a pronounced upward bias at later times (middle column in Fig.~\ref{Fig_windrose_3107}). Reorientation times exhibited only modest changes and at mid stationary phase perturbed cells showed slightly shorter characteristic times than static controls (Fig.~\ref{Fig_reorient_comparison_3107}B), consistent with a tendency toward more ballistic, stably oriented swimming once the perturbation had been accommodated.

The 168-hour delay scenario produced the smallest detectable change in HA3107 motility, with behaviour remaining close to the static reference. Mean vertical velocities of static and perturbed cultures remained similar throughout the remaining observation period. Directional distributions retained a strong upward anisotropy in both conditions, with only minor broadening of the angular spread in perturbed cells (right column in Fig.~\ref{Fig_windrose_3107}). Reorientation times overlapped within experimental uncertainty (Fig.~\ref{Fig_reorient_comparison_3107}C), indicating that late-stage forcing did not measurably disrupt gravitactic alignment.
Comparison with HA452 highlights strain-specific strategies for coping with hydrodynamic forcing in motile species. Under continuous early forcing, HA452 shifted toward slow, weakly directed motion with long reorientation times, while HA3107 retained substantial upward migration and orientational stability. Delayed forcing at 120 h or 168 h induced only temporary or very mild changes in HA3107 motility, whereas HA452 showed a clearer history dependence. These results support the view that HA3107 sits in a more mechanically robust sector of gravitactic phase space, where the balance between gravitational torque, viscous shear, and intrinsic noise allows the maintenance or rapid restoration of directed upward swimming even under sustained small-scale shear.

\FloatBarrier

\begin{figure}[H]
    \centering
    \includegraphics[width=0.7\textwidth]{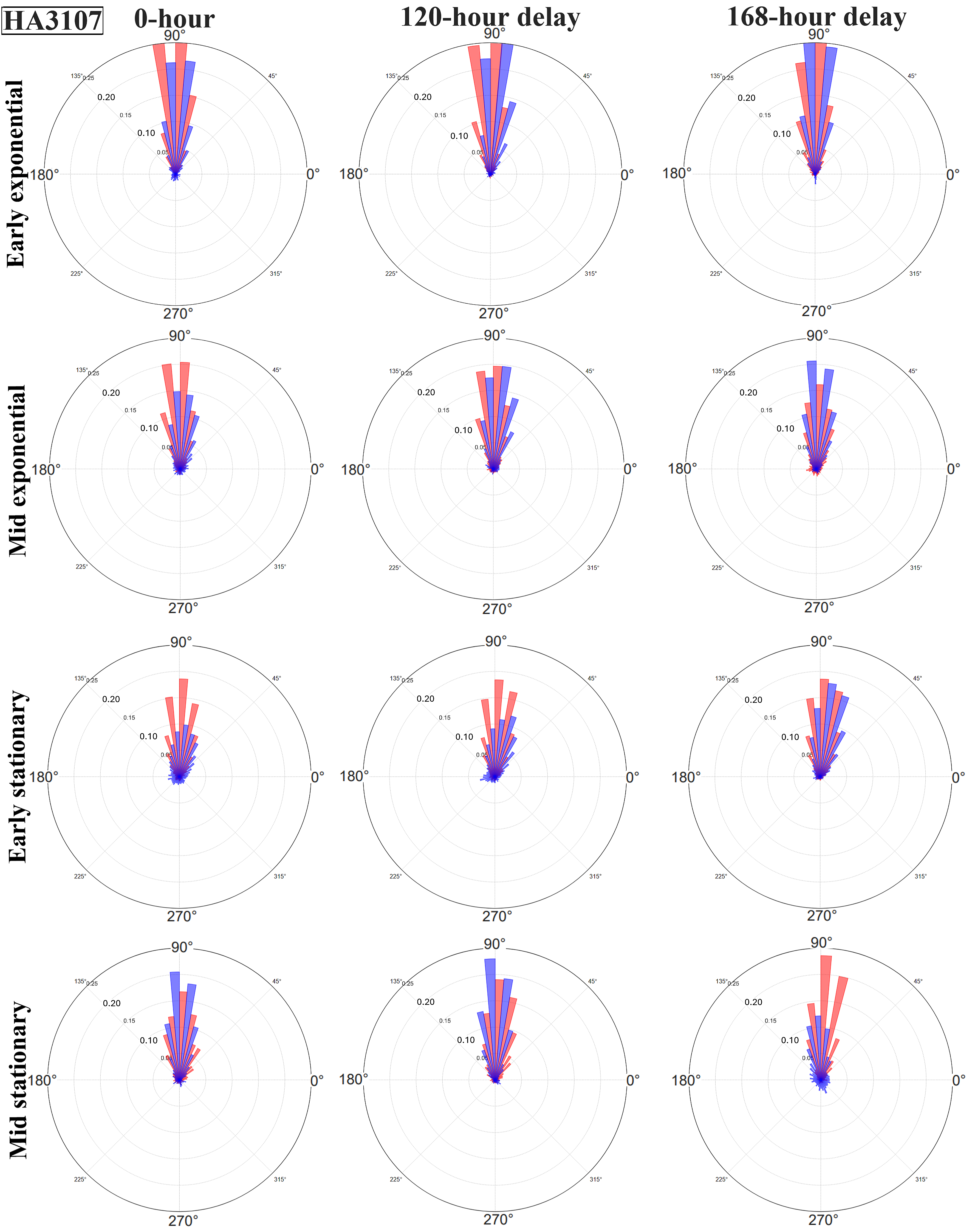}
\end{figure}
        \caption[Swimming direction distribution of \textit{H. akashiwo} (CCMP3107) cells under various perturbation scenarios]%
    {{\textbf{Distribution of swimming directions of HA3107 (standard scenarios).}
    Windrose plots illustrate the swimming direction of HA3107 cells for static cells (red) and perturbed cells (blue) across various perturbation scenarios.
    The length of each bar represents the relative frequency of cells swimming in that direction, with the frequencies indicated by labels on the concentric circles.
    Each row corresponds to a specific time point across all rounds, while each column represents a different perturbation scenario at various time points.
    HA3107 cells' swimming direction appears to be minimally affected by hydrodynamic perturbation.}}
    
    \label{Fig_windrose_3107}

Building on the standard scenarios, we next examine the reverse perturbation scenarios for HA452, in which cultures experienced orbital shaking immediately after inoculation but the flow was stopped at defined times and all subsequent growth proceeded under static conditions. We analysed three such histories that differed only in the duration of this initial forcing: continuous agitation throughout the experiment (0-hour delay), an early pulse terminated at 135 h, and a longer pulse terminated at 168 h. 

\FloatBarrier

\begin{figure}[H]
\centering
\includegraphics[width=1\columnwidth]{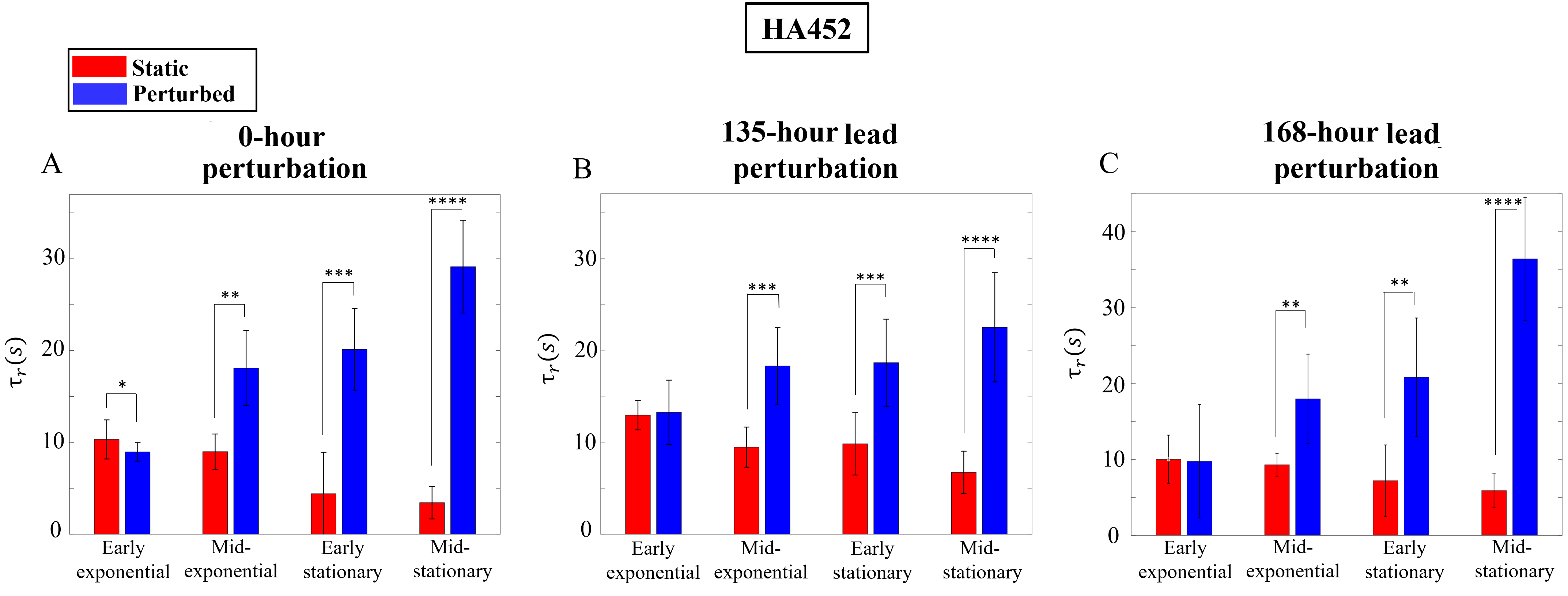}
\caption[Comparison of reorientation timescales  between static and perturbed \textit{H. akashiwo} (CCMP452) cells across various perturbation scenarios]{{\textbf{Comparison of reorientation timescales under reverse perturbation scenarios
for HA452. } {(A) Reorientation timescales of HA452 cells under the 0-hour perturbation scenario at four distinct time points: early exponential, mid-exponential, early stationary, and mid-stationary growth phases. (B) Reorientation timescales of cells under the 135-hour lead scenario. (C) Reorientation timescales of cells under a 168-hour lead scenario. Early hydrodynamic perturbation significantly affect the swimming stability of the cells. In all plots, the mean reorientation timescale of two biological and two technical replicates is presented, with standard deviation shown as error bars. Statistical significance is indicated as follows: (*) for $P < 0.05$, (**) for $P < 0.01$, (***) for $P < 0.001$, and (****) for $P < 0.0001$.}}}
\label{Fig_thau_452_backward}
\end{figure}

Early forcing that was stopped near the onset of exponential growth already produced a long-lived motility signature. In the 135-hour early scenario, cultures were agitated from inoculation until 135~h and then returned to static conditions. Despite this return, perturbed cells maintained strongly reduced upward swimming velocities compared with static controls at all subsequent time points (Fig.~\ref{vertical_452_backward}B). During early and mid exponential phases the mean vertical speed of history-perturbed cells remained close to zero or even slightly negative. The corresponding wind-rose plots show that static cultures rapidly developed a narrow upward lobe, while the history-perturbed population displayed a much broader distribution with enhanced lateral components and, at some stages, almost isotropic swimming (middle column in Fig.~\ref{Fig_windrose_backward_452}). Characteristic reorientation times were consistently higher for such populations than in controls (Fig.~\ref{Fig_thau_452_backward}B), indicating slower relaxation toward the stable upward orientation even though all cultures experienced the same quiescent fluid environment during measurement. Thus, a few days of early shear exposure were sufficient to reprogram HA452 motility into a less directed, less stable state that persisted long after the mechanical stimulus was removed.

Extending the lead forcing window reinforced this memory. In the 168-hour lead scenario, cultures were agitated from inoculation until late exponential or early stationary phase and then kept static. At the later observation times the vertical velocities of the perturbed cells again remained much lower than those of the static reference (Fig.~\ref{vertical_452_backward}C). Static cultures showed the familiar increase in upward swimming across the growth curve, whereas perturbed cells failed to reach comparable speeds and in some phases hovered around weakly upward or near-zero motion. Directional statistics confirmed that the upward anisotropy of static controls was only partially recovered in the history-perturbed population, which retained broader lobes and an over-representation of oblique trajectories (right column in Fig.~\ref{Fig_windrose_backward_452}). The characteristic reorientation time $\tau_c$ remained significantly elevated for the history-perturbed cultures (Fig.~\ref{Fig_thau_452_backward}C), consistent with a motility state in which cells reorient slowly and spend longer times misaligned with the vertical.

Comparison of the 135-hour and 168-hour lead scenarios with the continuously forced 0-hour delay case emphasizes that the key factor is the presence of hydrodynamic perturbation immediately after inoculation rather than its duration. In all three histories, populations which experienced shear from the very beginning adopted reduced upward swimming, broadened directional distributions, and increased reorientation times, even when the flow was later switched off and the fluid environment became comparable to that of the controls. The similarity between the short (135~h) and long (168~h) intervals, along with continuously forced cases suggests that the first few days after inoculation define a sensitive window during which mechanical cues can lock motile swimmers into an altered motility state that is effectively irreversible on experimental time scales.

\begin{figure}[H]
    \centering
    \includegraphics[width=1\textwidth]{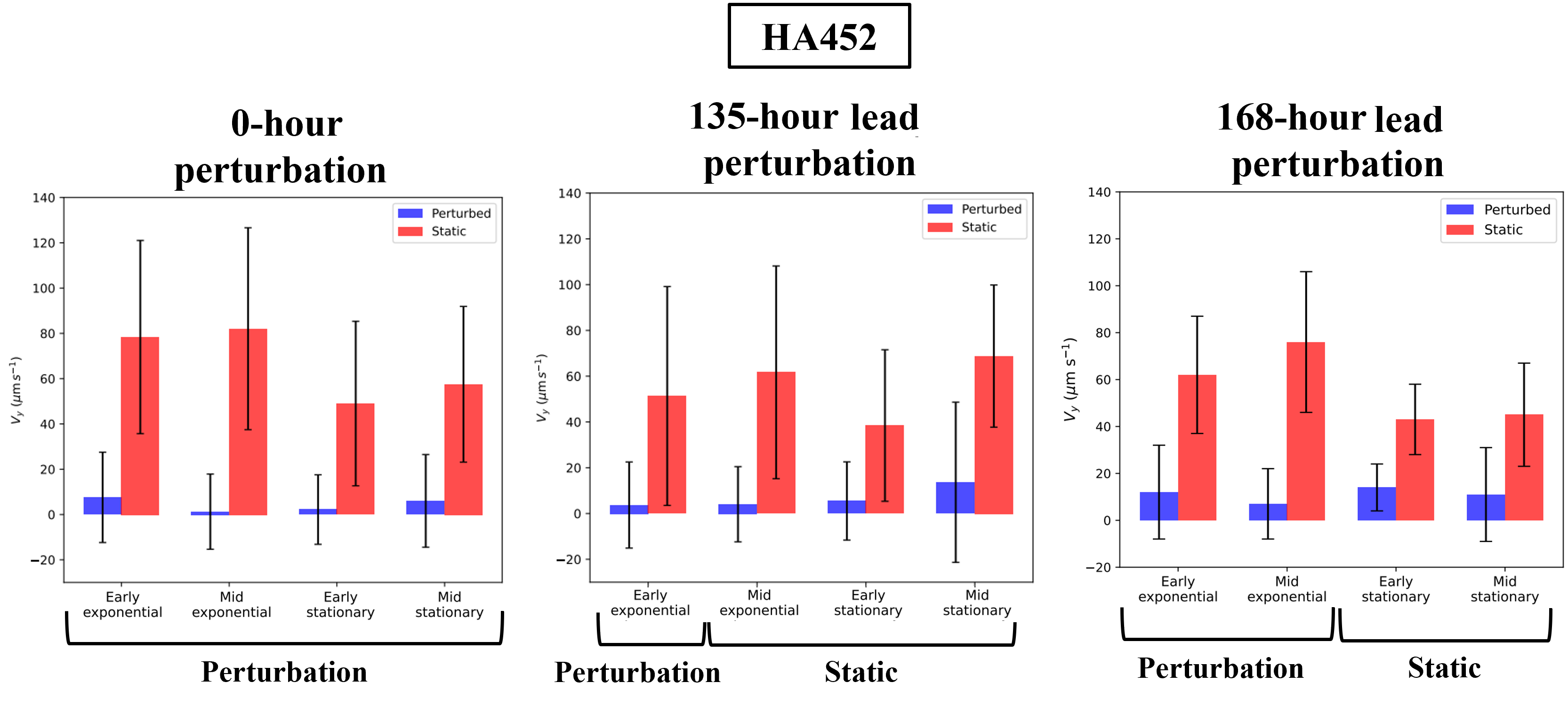}
\end{figure}
\captionof{figure}{\textbf{Emergent vertical swimming velocity of HA452 under reverse hydrodynamic scenarios.} Mean vertical velocities (upward positive) are shown for perturbed cultures (blue) and static controls (red) across the four growth phases and perturbation scenarios. Cells exposed to pertubations from the very beginning of their life in culture adopted a motile phenotype, revealed by significant reduction of up swimming. Values represent means from two independent biological replicates with two technical replicates per time point.}
\label{vertical_452_backward}

This irreversibility is consistent with the idea that early hydrodynamic forcing triggers deeper cellular reprogramming rather than a purely mechanical, instantaneous response. Several mechanisms are plausible. The lipid measurements in HA452 show that early and sustained forcing promotes strong accumulation \cite{kakavand2025hydrodynamic} and probably spatial reorganization of cytoplasmic lipid droplets, which modify the gravitational torque. In parallel, mechanical cues can modulate gene expression and signalling pathways that regulate flagellar waveform, beat asymmetry, and mechanosensory feedback. Such changes can alter the effective gravitactic parameter and noise level that control the balance between gravitational and viscous torques. Once established, these internal reorganizations will persist even after the external shear has ceased to act, thereby maintaining long reorientation times and weakly directed vertical migration. In this view, the reverse scenarios reveal a form of hydrodynamic "memory" in motile microorganisms, where early hydrodynamic history is written into cellular architecture and motility control, and continues to shape swimming behaviour long after the flow itself has been removed.

\begin{figure}[H]
    \centering
    \includegraphics[width=0.8\textwidth]{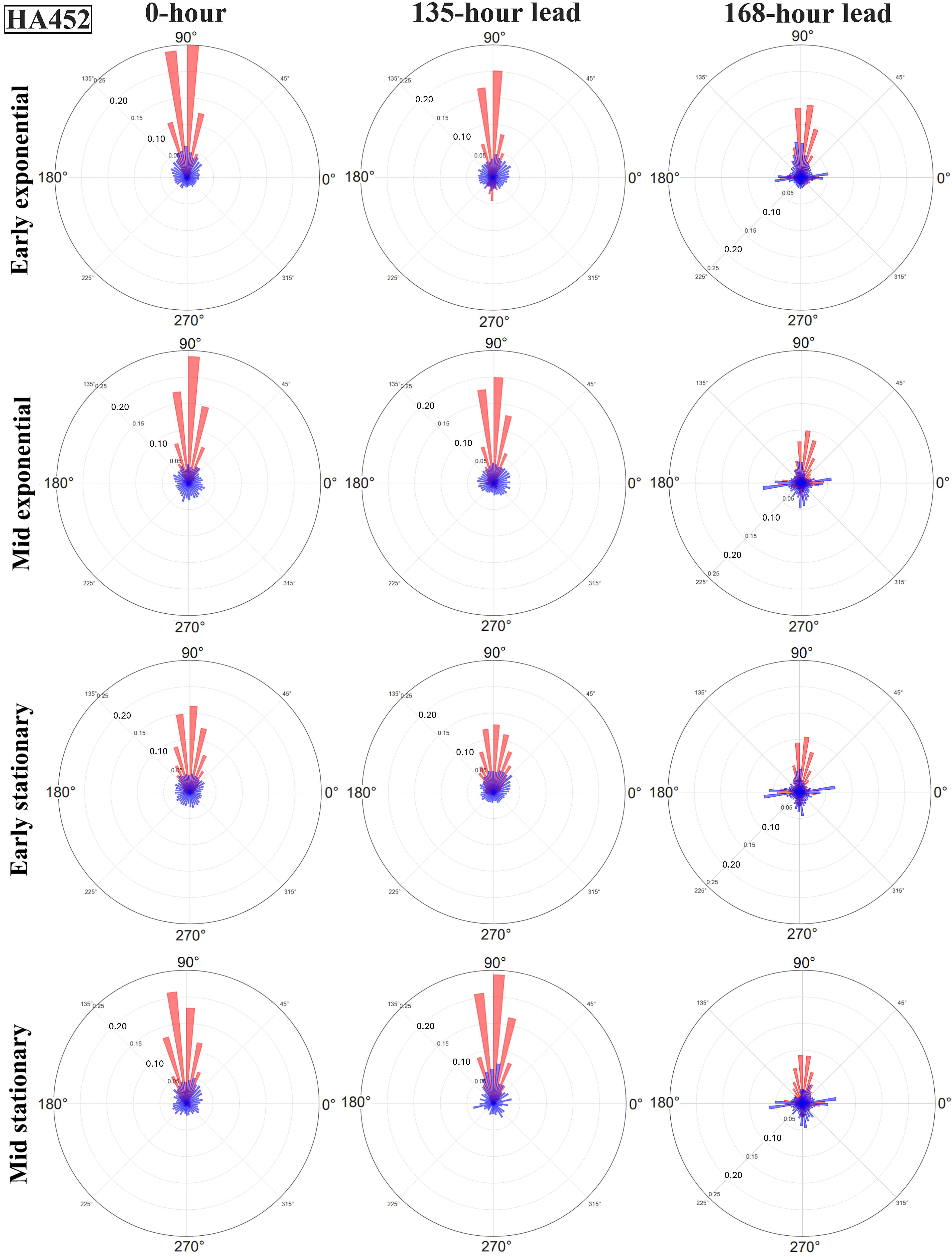}
\end{figure}
\caption[Swimming direction distribution of \textit{H. akashiwo} (CCMP3107) cells under various perturbation scenarios]%
{{\textbf{Distribution of swimming directions of HA452 cells over different reverse perturbation scenarios.}
    Windrose plots illustrate the swimming direction of HA452 cells for static cells (red) and perturbed cells (blue) across various perturbation scenarios. Each row corresponds to a specific time point across all rounds, while each column represents a different perturbation scenario at various time points.
   HA452 cells that experienced shear immediately after inoculation adopted an isotropic swimming phenotype.}}
    
    \label{Fig_windrose_backward_452}

\begin{figure}
\centering
\includegraphics[width=1\columnwidth]{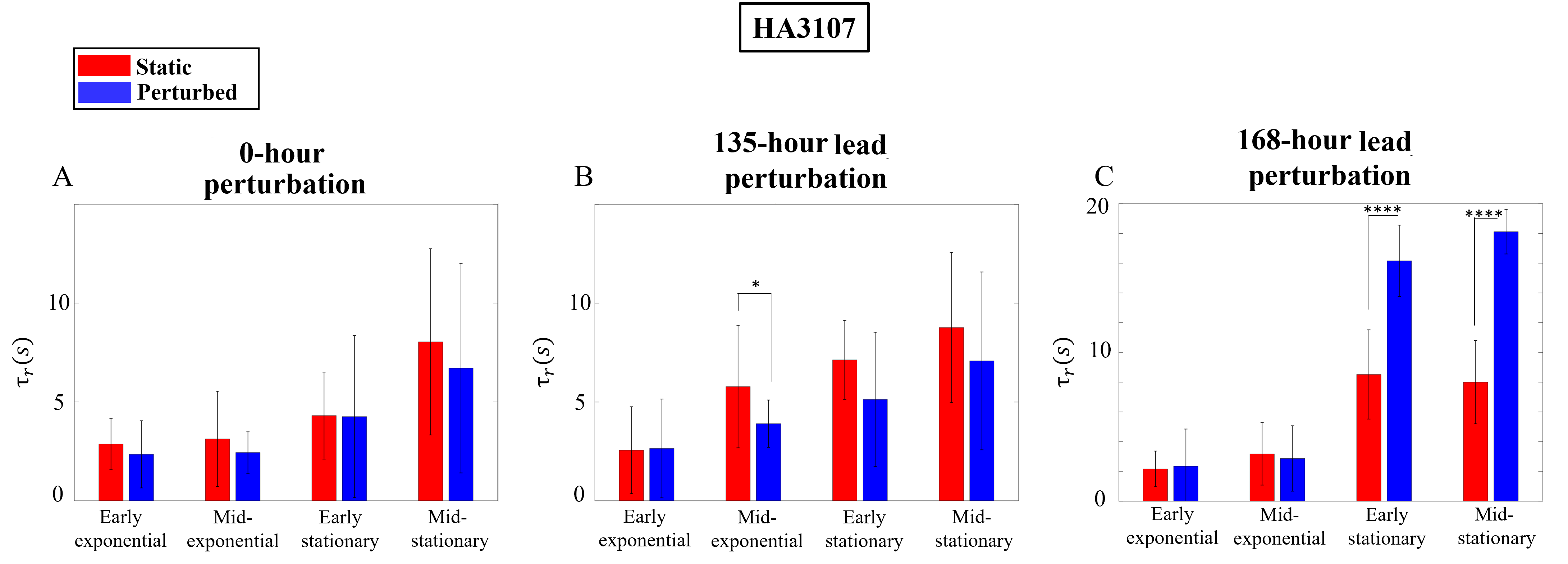}
\caption[Comparison of reorientation timescales between static and perturbed \textit{H. akashiwo} (CCMP3107) cells across various backward perturbation scenarios]{{\textbf{Comparison of reorientation timescales under reverse perturbation scenarios for HA3107.} {(A) Reorientation timescales of HA3107 cells under the 0-hour perturbation scenario at four distinct time points: early exponential, mid-exponential, early stationary, and mid-stationary growth phases. Early and continuous hydrodynamic perturbation did not significantly affect the swimming stability of the cells. (B) Reorientation timescales of cells under the 135-hour lead scenario. remained closely aligned with those of the never-perturbed controls at all later stages. (C) Reorientation timescales of cells under the 168-hour lead scenario. immediately after the transition to static conditions, mean vertical velocities of history-perturbed cells dropped markedly across subsequent phases. In all plots, the mean reorientation timescale of two biological and two technical replicates is presented, with standard deviation shown as error bars. Statistical significance is indicated as follows: (*) for $P < 0.05$, (**) for $P < 0.01$, (***) for $P < 0.001$, and (****) for $P < 0.0001$.}}}
\label{Fig_thau_3107_backward}
\end{figure}

For HA3107, reverse scenarios confirm the overall robustness inferred from the standard experiments but also exposed a striking sensitivity to the duration of early forcing. After a short duration of agitation (135-hour early), vertical velocities, directional distributions, and characteristic reorientation times of the perturbed population remained closely aligned with those of the never-perturbed controls at all later stages (Figs.~\ref{vertical_3107_backward}B, \ref{Fig_windrose_backward_3107}, \ref{Fig_thau_3107_backward}B). Upward migration and gravitactic stability were therefore fully restored once the flow was removed, indicating that brief early exposure does not imprint a lasting ecological memory on this strain.

\FloatBarrier

By contrast, extending the early perturbation to 168 h qualitatively changed the outcome. While cultures were still on the shaker, HA3107 maintained strong upward swimming and short reorientation times, comparable to static controls (Figs.~\ref{vertical_3107_backward}C and \ref{Fig_thau_3107_backward}C). However, immediately after the transition to static conditions, mean vertical velocities of perturbed cells dropped markedly across subsequent phases and remained well below control values. Swimming direction distribution plots show that the upward swimming collapsed into a much broader distribution with enhanced oblique and lateral trajectories (right column in Fig.~\ref{Fig_windrose_backward_3107}), and the characteristic reorientation time increased significantly, consistent with slower realignment toward the stable upward orientation (Fig.~\ref{Fig_thau_3107_backward}C). This delayed collapse suggests that prolonged acclimation to shear creates a dependence of the motility machinery on persistent mechanical input; once the shear is removed, cells enter a mechanically "over-relaxed" state with weakened gravitactic control. Mechanistically, such behaviour could arise from long-term mechanotransduction and transcriptional responses that tune flagellar beating, mechanosensory feedback, and internal mass distribution to the perturbed regime. When the external forcing is suddenly switched off, those slow regulatory adjustments become maladaptive, leading to sustained reductions in swimming speed and orientational stability despite the apparently benign static environment.

\begin{figure}[H]
    \centering
    \includegraphics[width=1\textwidth]{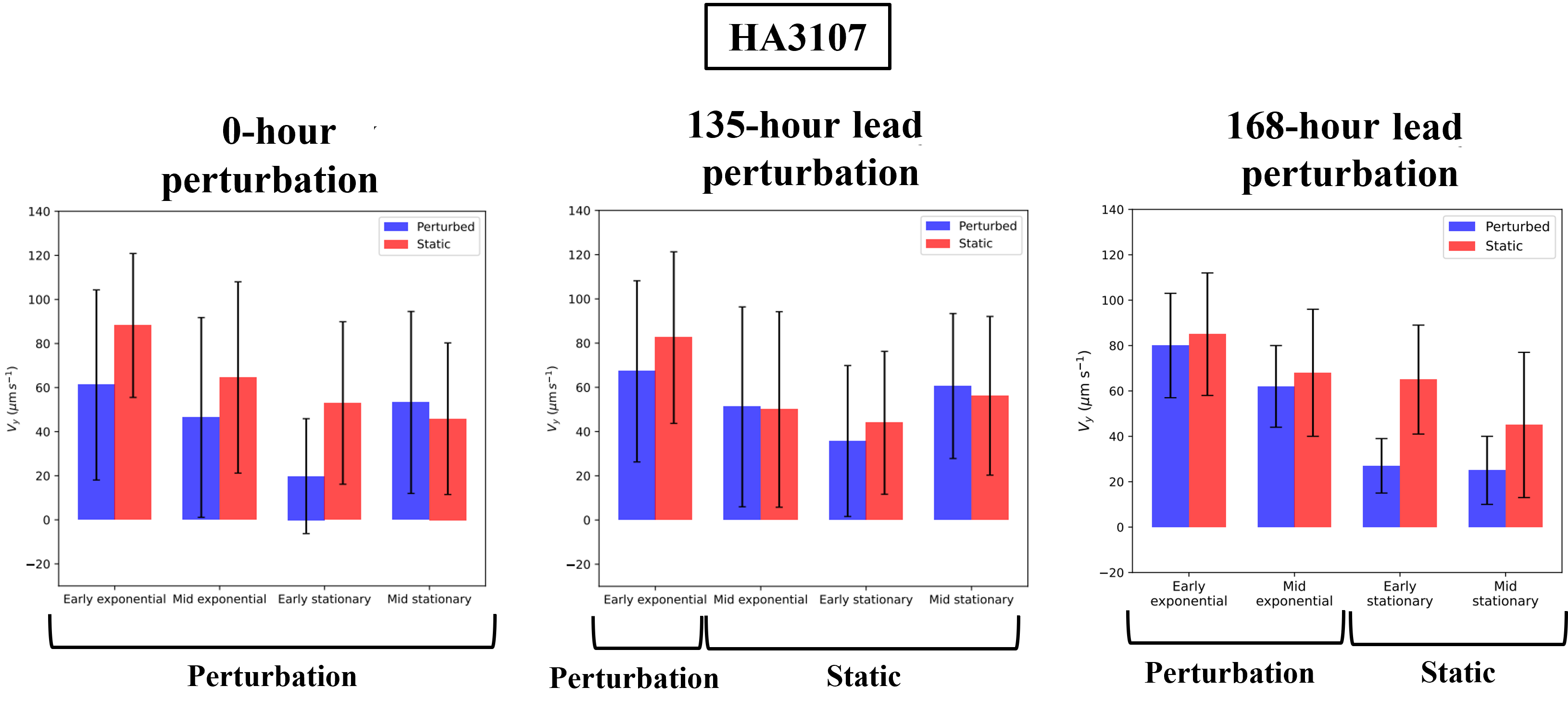}
\end{figure}
\captionof{figure}{\textbf{Vertical swimming velocity of HA3107 under different reverse perturbation scenarios.} Mean vertical velocities (upward positive) are shown for perturbed cultures (blue) and static controls (red) across the four growth phases and perturbation scenarios. In contrast to strain HA452, perturbation timing did not significantly alter the vertical swimming behavior of HA3107 cells, with velocities remaining comparable to controls throughout the experiments. Values represent means from two independent biological replicates with two technical replicates per time point.}
\label{vertical_3107_backward}

\subsection*{Hydrodynamic perturbation regulates lipid droplet accumulation in a growth‐stage–dependent manner}

Cytoplasmic lipid accumulation in HA452 was quantified across all forward hydrodynamic-perturbation scenarios to examine how growth-stage dependent mechanical forcing reshapes cellular energy storage and internal mass distribution \cite{Sengupta2022Active,kakavand2025hydrodynamic}. Lipid droplets were stained and imaged at the single-cell level, and two complementary metrics were extracted: the normalized lipid area, which reports the projected lipid-occupied fraction of the cytoplasm, and the total lipid droplet volume per cell, which integrates the three-dimensional lipid load \cite{Sengupta2022Active}. For each scenario, cells were sampled immediately before the onset of hydrodynamic perturbation, within the first 24–48~h after the transition, and at later times when the consequences of the altered mechanical environment had fully developed, consistent with earlier observations that pronounced differences in lipid content emerge over this interval \cite{Sengupta2022Active,kakavand2025hydrodynamic}. This design allows the temporal structure and magnitude of the lipid response to be compared across different perturbation timings along the growth curve.

In the 0-hour delay scenario, where hydrodynamic perturbation was applied from inoculation and maintained throughout the experiment, HA452 cells displayed a pronounced and sustained enhancement of lipid content compared with static controls. Both normalized lipid area and total lipid volume per cell increased rapidly after the onset of exponential growth and continued to rise into stationary phase, reaching the highest values observed in any forward scenario. The distributions of single-cell lipid loads were broad, with large cell-to-cell variability superimposed on the elevated mean, indicating heterogeneous but systematically upregulated lipid storage in the perturbed population \cite{Sengupta2022Active}. These data show that continuous exposure to hydrodynamic perturbation from the earliest growth stages strongly promotes the diversion of carbon into cytoplasmic lipid droplets in HA452.

When hydrodynamic perturbation was introduced after a period of initial static growth, the amplitude of the lipid response was reduced but still evident. In the 120-hour delay scenario, pre-transition lipid levels were indistinguishable between control and experimental cultures by construction. Following the switch to orbital shaking, both normalized lipid area and total lipid volume per cell increased relative to the static controls, but the enhancement developed more gradually and reached values lower than those attained under the 0-hour delay. A similar behaviour was observed in the 168-hour delay scenario, where the transition occurred later along the exponential phase. In both delayed-onset regimes, hydrodynamic perturbation thus induced additional lipid accumulation on top of the baseline growth-associated increase, yet the effect size did not match the strong lipid loading achieved under continuous early forcing.

In the 348-hour delay scenario, perturbation was applied only after cultures had already reached stationary phase. At this point, lipid content had already risen in both static and to-be-perturbed populations as part of the stationary-phase programme. Subsequent exposure to hydrodynamic perturbation caused, at most, modest further changes in normalized lipid area and total lipid volume, and the trajectories of perturbed and static cultures remained relatively close. This indicates that once HA452 populations have completed the major phase of growth-associated lipid accumulation, late mechanical forcing has limited capacity to further remodel their cytoplasmic lipid stores \cite{kakavand2025hydrodynamic}.

Across the forward scenarios, these observations establish that the magnitude and variability of lipid droplet accumulation in HA452 depend strongly on the timing of hydrodynamic perturbation. Mechanical forcing present from inoculation elicits the largest and most heterogeneous increase in cytoplasmic lipid content, whereas perturbations imposed later along the growth curve generate progressively weaker lipid responses and have little additional effect when applied after the onset of stationary phase. Because lipid droplets are denser than the surrounding cytoplasm and typically form non-uniformly within the cell body, this history-dependent lipid remodelling is expected to shift the center of mass relative to the geometric center, altering cellular bottom-heaviness and the gravitational torque acting on swimming cells \cite{Sengupta2022Active}. The forward lipid data for HA452 therefore provide a mechanistic link between hydrodynamic history, metabolic reallocation into storage lipids, and the mass-distribution changes that can feed back on orientational stability and motility, as explored in the motility and discussion sections \cite{Sengupta2022Active,kakavand2025hydrodynamic}.

\section*{Summary and discussion}

Our results show that small-scale hydrodynamic forcing does not have a single, fixed effect on motile phytoplankton. The same level of shear can either support, neutralise or damage population performance depending on when it is experienced along the growth trajectory, and on the prior hydrodynamic history of the cells. At the cellular level, this history is written into the architecture of cytoplasmic lipids and, through them, into the torque balance that underlies gravitactic motility, consistent with earlier observations on \textit{Heterosigma akashiwo} and related microalgae \cite{Sengupta2017Phytoplankton, Sengupta2022Active}. 

Together, these findings identify hydrodynamic history as a hidden control parameter for both fitness and motility in \textit{H.~akashiwo}. The first key point is that the impact of shear is strongly path-dependent. If hydrodynamic forcing is present while populations are still establishing, cells can integrate this cue into their developmental programme: growth remains fast, carrying capacity is preserved, and photophysiology is only modestly perturbed (data not shown here). Once exponential growth is underway, the same forcing produces a qualitatively different response: carrying capacity drops, photophysiological indices deteriorate and survival is shortened. The difference cannot be explained by intensity of perturbation, which is held constant, but by the timing of the perturbation relative to the internal state of the population. This implies the existence of a finite “plasticity window” early in the growth curve, during which the metabolic and regulatory networks which control resource allocation, stress responses and cell architecture remain reconfigurable in response to mechanical cues (cf. turbulence-induced plasticity in motile phytoplankton reported in \cite{Sengupta2017Phytoplankton, Sengupta2022Active}. Beyond this window, the cost of reconfiguring an already committed programme appears to outweigh potential benefits, so shear is expressed mainly as damage.

The reverse scenarios confirm that this response is not only time-dependent but also hysteretic. Populations which experienced shear early and were then returned to quiescent conditions do not simply relax back to the static phenotype. Instead, they retain altered growth and photophysiological properties, as well as modified motility, long after the flow has been removed. This ecological memory is asymmetric between strains: HA452 pays a relatively small fitness cost under early forcing but shows strong and persistent motility changes, whereas HA3107 maintains robust gravitaxis over a wider range of histories but suffers larger reductions in biomass and photophysiology when shear is imposed or removed late. In other words, the two strains occupy different points on a trade-off between mechanical tolerance of growth and mechanical robustness of motility, echoing strain-specific trade-offs reported previously for mechanically and light-driven stress responses in phytoplankton.

\newpage

\begin{figure}[H]
    \centering
    \includegraphics[width=0.7\textwidth]{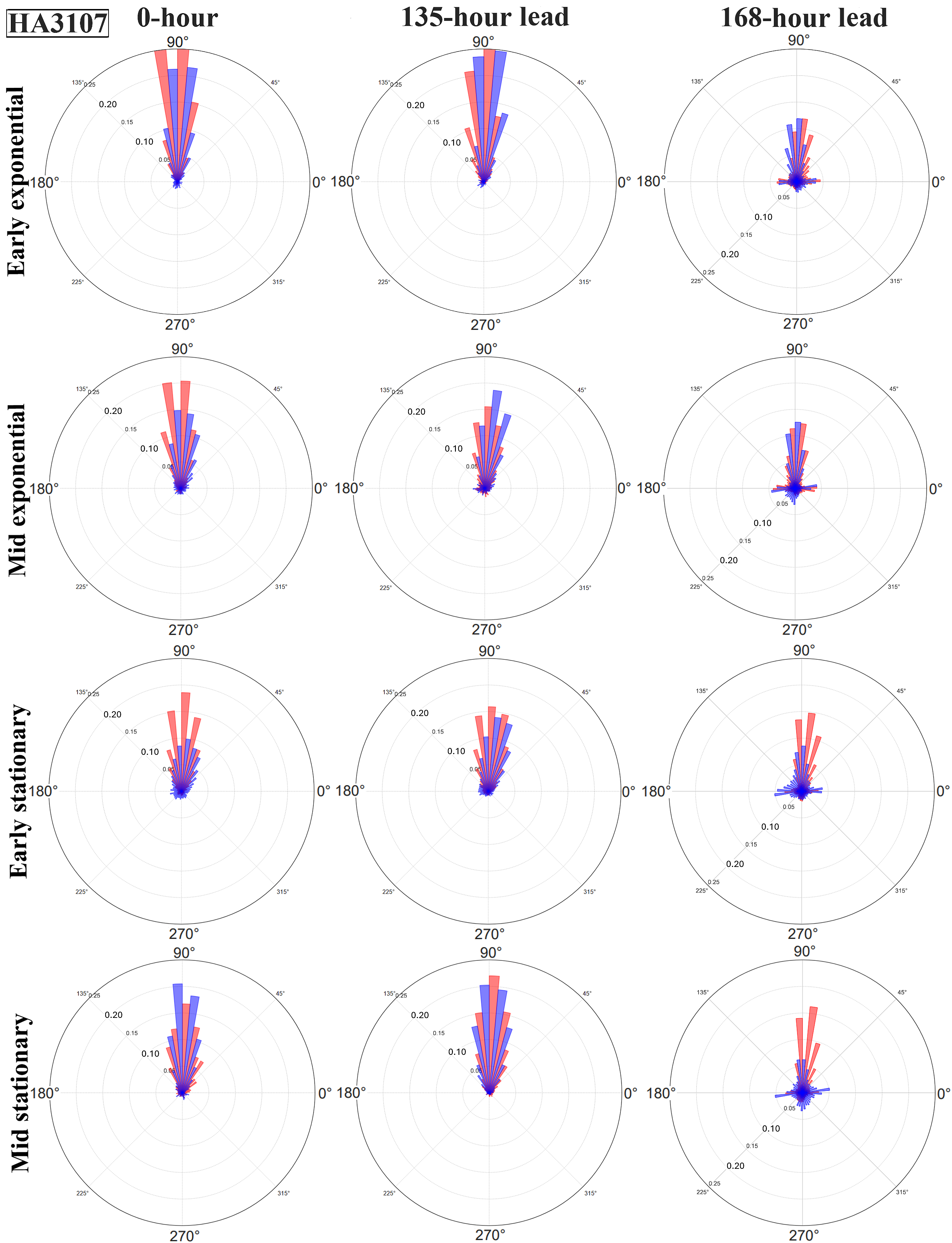}
\end{figure}
     \caption[Swimming direction distribution of \textit{H. akashiwo} (CCMP3107) cells under reverse perturbation scenarios]%
    {{\textbf{Distribution of swimming directions of HA3107 cells across reverse perturbation scenarios.}
    Windrose plots illustrate the swimming direction of HA3107 cells for static cells (red) and perturbed cells (blue) across various perturbation scenarios.
    The length of each bar represents the relative frequency of cells swimming in that direction, with the frequencies indicated by labels on the concentric circles.
    Each row corresponds to a specific time point across all rounds, while each column represents the perturbation scenarios at various time points.
    HA3107 swimming direction appears to be minimally affected by hydrodynamic cues.}}
    
    \label{Fig_windrose_backward_3107}

\begin{figure}
\centering
\includegraphics[width=1\columnwidth]{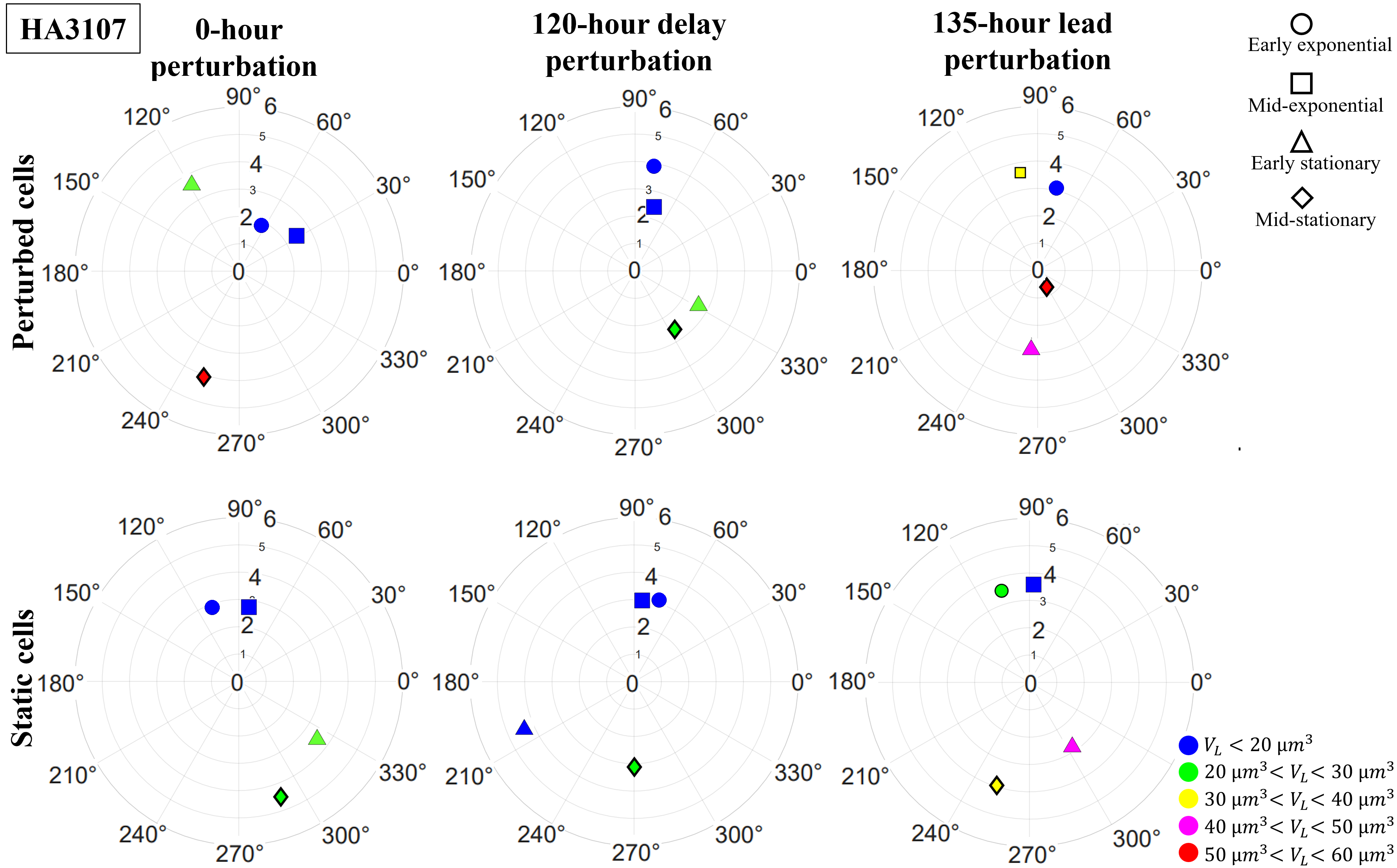}
\caption[Lipid droplet volume and spatial distribution relative to the cell's centroid across different growth phases for different perturbation scenarios of HA3107 cells]{\textbf{Effective volume and spatial distribution of lipid droplets in HA3107 relative to the cell's centroid.} The equivalent lipid droplet (LD) volume and its relative position to the cell’s centroid (center of the plot) are presented for the 0-hour, 120-hour delay, and 135-hour lead perturbation scenarios in the first, second, and third columns, respectively, with perturbed cells shown in the first row and static cells in the second row. The LD volume is represented by a color gradient, and each growth phase is distinguished by a specific geometric symbol. The data were obtained by analyzing 40 cells per time point. The radial distance from the center is in $\mu$m.}
\label{Fig_lipid_location_3107}
\end{figure}
\begin{figure}
\centering
\includegraphics[width=1\columnwidth]{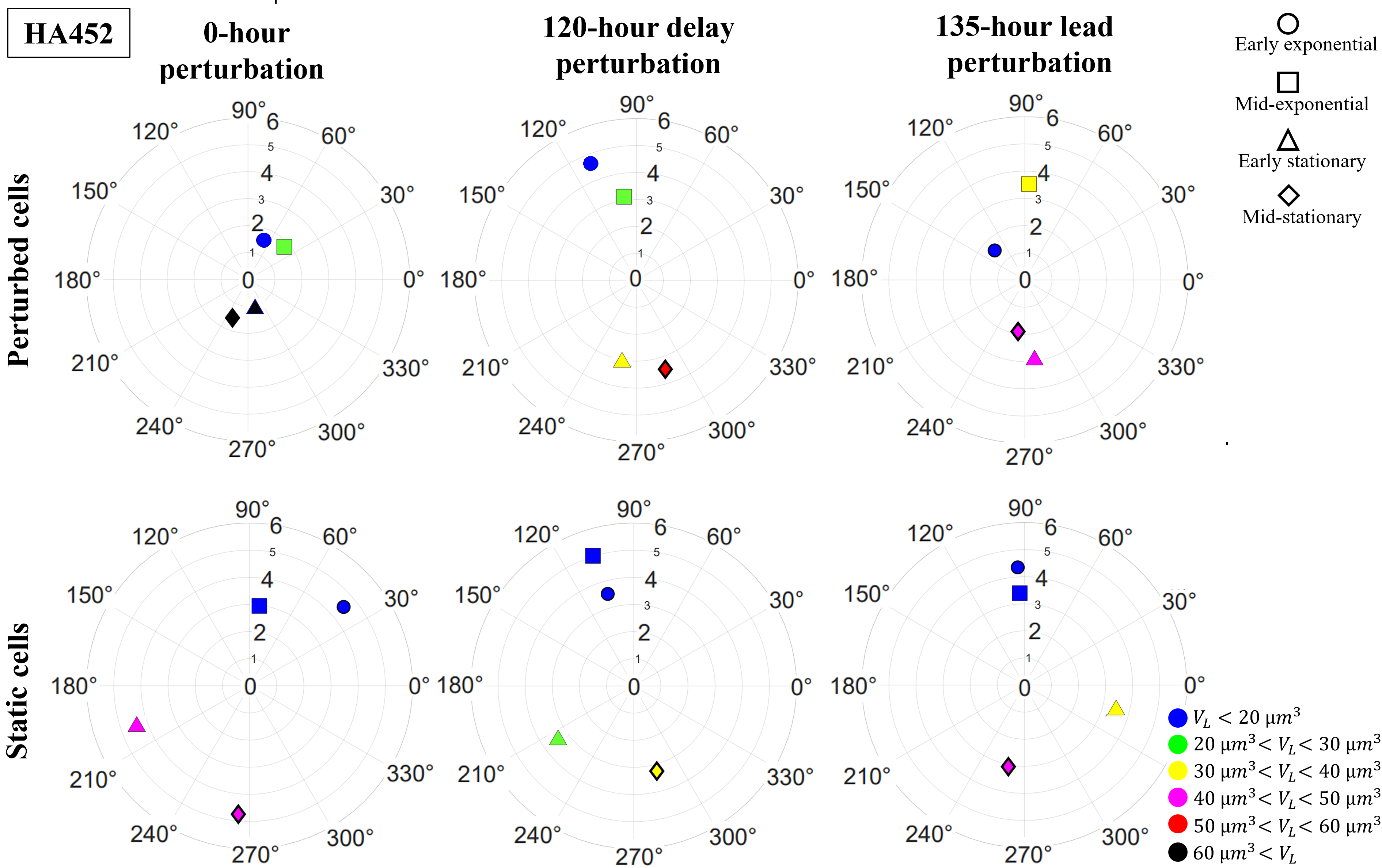}
\caption[Lipid droplet volume and spatial distribution relative to the cell's centroid across different growth phases for different perturbation scenarios of HA452 cells]{\textbf{Effective volume and spatial distribution of lipid droplets in HA452 relative to the cell's centroid.} The effective lipid droplet (LD) volume and its relative position to the cell’s centroid (center of the plot) for the 0-hour, 120-hour delay, and 135-hour lead perturbation scenarios in the first, second, and third columns, respectively, with perturbed cells shown in the first row and static cells in the second row. The LD volume is represented by a color gradient, and each growth phase is distinguished by a specific geometric symbol. The data were obtained by analyzing 40 cells per time point. The radial distance from the center is in $\mu$m.}
\label{Fig_lipid_location_452}
\end{figure}

To interpret the motility data mechanistically, we build directly on the motility model developed in our earlier work \cite{Sengupta2022Active}. In that framework, the complex ensemble of intracellular lipid droplets is coarse-grained into a single “effective lipid droplet”, defined as the volume-weighted centroid of all lipid droplets within the cell. The position of this effective droplet relative to the geometric centre of the cell sets both the magnitude and the direction of the cell density offset that generates the gravitational torque. Its radial distance controls the lever arm of the weight force, while its polar angle, measured with respect to the flagellar axis (taken as $90^{\circ}$ in the polar representation), fixes the orientation of the torque. This effective droplet is therefore the key mechanical variable that links metabolic lipid allocation to the torque balance governing gravitactic alignment. In this context, Figs.~\ref{Fig_lipid_location_452} and~\ref{Fig_lipid_location_3107} report, for each history, the effective lipid-droplet volume and its position in polar coordinates, with the origin at the geometrical centre of the cell and the flagellar pole at $90^{\circ}$. These polar plots thus provide a direct map from intracellular lipid architecture to the parameters of the motility model.

\FloatBarrier

The second key point is that the observed history-dependence has a clear mechanical basis when expressed in terms of this effective droplet. Gravitactic alignment in motile microalgae is governed by a balance between gravitational and viscous torques, controlled by the offset between the cell’s centre of mass and its geometric and hydrodynamic centres. In this picture, the effective lipid droplet acts as a mobile, dense inclusion that can shift the centre of mass and thereby tune the effective gravitational torque. Our lipid-location maps show that hydrodynamic history modifies not only the total lipid volume but, crucially, its spatial distribution within the cell. When shear is applied during the early plasticity window, lipid droplets in HA452 tend to grow and relocate towards the geometric centre, so that a large internal mass is concentrated near the centroid. This configuration increases cellular density and energy storage but reduces the moment of the gravitational force. In torque-balance terms, histories with early forcing push the population toward a region of parameter space with high lipid volume but small gravitational offset, weakening the restoring torque and allowing rotational noise and hydrodynamic disturbances to dominate. The corresponding polar plots show isotropic direction distributions and long reorientation times: a mechanically reprogrammed motility with weak effective gravitaxis.

By contrast, in scenarios where perturbations are absent or imposed only after exponential growth has been established, lipid droplets in both strains follow trajectories that maintain a substantial offset from the centroid. Here the product of lipid volume and offset remains large, so the gravitational torque stays strong and cells preserve a clear upward swimming bias with short reorientation times. HA3107, in particular, tends to keep lipid droplets in configurations that sustain bottom-heaviness over a broad range of histories; in the conceptual phase space defined by lipid volume and offset, this strain occupies a region where the gravitactic torque is robust to mechanical perturbation. In this model, the strain differences emerge not from qualitatively different physics but from distinct trajectories in a shared mechanical phase space: hydrodynamic history moves each strain along a different path in the (lipid volume, lipid offset) plane, and their motility phenotypes correspond to different regions of this plane.

This torque-based view also clarifies why motility and fitness are not trivially correlated. For HA452, early shear drives cells into a high-lipid, weak-torque regime: growth and lipid storage remain favourable, but gravitaxis is sacrificed, yielding isotropic motility that would profoundly change how cells sample vertical gradients in a water column. For HA3107, the same class of histories keeps the population in a strong-torque regime: gravitaxis is preserved at the expense of greater sensitivity of growth and photophysiology to late perturbations. Thus, mechanical history does not simply “stress’’ cells; it selects among alternative mechanical phenotypes that differ in how they trade off directional motility against metabolic performance.

In summary, the data argue for a compact mechanistic picture. Hydrodynamic cues experienced within a finite early window reshape metabolic allocation and lipid trafficking, which in turn reposition cytoplasmic lipids and shift the centre-of-mass relative to the cell body. This remodelling moves populations between distinct regions of a gravitactic phase space defined by effective lipid-droplet volume and offset, altering the balance of gravitational and viscous torques and producing history-dependent motility states. The same histories also leave a lasting imprint on growth and photophysiology, with strain-specific trade-offs between mechanical robustness of fitness and motility. Recognising hydrodynamic history as a control parameter in this coupled mechanical–physiological system is essential for predicting the response of motile phytoplankton to realistic, time-varying flows.

\newpage

\section*{Materials and Methods}

\label{methods}

\subsection*{Experiments under static conditions}

Static culture conditions (defined as conditions where no hydrodynamic agitation or external physical forcing is applied throughout the entire growth period) for two \textit{H.~akashiwo} strains (CCMP452 and CCMP3107; hereafter HA452 and HA3107) followed established protocols adapted from \cite{Sengupta2022Active, Guillard1975}. Cells were maintained in sterile 50~mL glass tubes containing f/2 medium without silicate (f/2--Si) prepared in artificial seawater (ASW). ASW was prepared in-house by dissolving 36~g of commercial sea salt (Instant Ocean\textsuperscript{\textregistered}, Aqua Systems) per litre of Milli-Q\textsuperscript{\textregistered} water (final salinity 35--36~PSU; pH 8.0~$\pm$~0.1). The salt solution was stirred at 500--600~rpm for 1.5--2~h on a magnetic stirrer (IKA\textsuperscript{\textregistered} RCT basic) and then sterile-filtered through a 0.2~$\mu$m polyethersulfone (PES) bottle-top membrane (VWR\textsuperscript{\textregistered}). Nutrient enrichment was performed aseptically in a UV-sterilized laminar-flow hood by adding nitrate and phosphate stock solutions to reach the canonical f/2(–Si) macronutrient concentrations (882~$\mu$M nitrate as NaNO$_3$ and 36~$\mu$M phosphate as NaH$_2$PO$_4\cdot$H$_2$O), together with trace metals and vitamins at standard f/2 levels \cite{Guillard1975} (Bigelow Laboratory nutrient kits). Silicic acid was omitted to match the physiology of non-silicifying \textit{H.~akashiwo}. For laboratories preferring a fully defined ASW at $\sim$35~PSU, an equivalent per-litre composition is: NaCl 24.53~g, MgCl$_2\cdot$6H$_2$O 10.78~g, Na$_2$SO$_4$ 4.09~g, CaCl$_2\cdot$2H$_2$O 1.47~g, KCl 0.66~g, NaHCO$_3$ 0.172~g, KBr 0.100~g, H$_3$BO$_3$ 0.026~g, SrCl$_2\cdot$6H$_2$O 0.024~g, and NaF 0.003~g.

All handling and transfers were carried out under aseptic conditions inside a Class~II laminar-flow cabinet. The work area was disinfected with 70\% ethanol before and after each session, glassware and pipette tips were autoclaved prior to use, and all liquid manipulations were performed with sterile serological pipettes to minimise contamination. Operators wore sterile gloves and re-sanitised them with ethanol during culture handling.

For each strain, two independent biological replicates were propagated from exponentially growing mother cultures. Briefly, 2~mL of well-mixed cell suspension was aseptically sampled from the upper 0.5~cm of the mother culture and inoculated into 25~mL of fresh f/2--Si medium in separate tubes, corresponding to a medium:inoculum ratio of approximately 25{:}2. Each biological replicate was maintained in its own tube to avoid cross-contamination, and strains were handled in parallel but kept physically separate throughout. Cultures were incubated under a 14{:}10~h light:dark cycle at 22~$^\circ$C in a temperature-controlled chamber with white LED illumination (peak wavelength $\sim$535~nm) at an intensity of $1.35~\mathrm{mW\,cm^{-2}}$ (approximately $60~\mu\mathrm{mol\ photons\,m^{-2}\,s^{-1}}$) during the light phase. Lights were on from \texttt{04{:}00} to \texttt{18{:}00} and off from \texttt{18{:}00} to \texttt{04{:}00}. Culture propagation followed this inoculation scheme, with transfers performed from late exponential-phase mother cultures every 2--3~weeks to maintain active growth. All static experiments and routine culturing were conducted without imposed hydrodynamic agitation. Unless otherwise specified, experimental observations were based on cultures sampled between \texttt{10{:}00} and \texttt{16{:}00} during the light phase, at clearly defined growth stages described in the relevant sections.

\subsection*{Experimental design under dynamic conditions}
\label{sec:dynamic_design}

Here, dynamic conditions denote treatments in which \textit{H.~akashiwo} cultures were subjected to a controlled hydrodynamic perturbation on an orbital shaker. All cultures (dynamic and static) were propagated and handled exactly as described above under sterile conditions. Building on preliminary observations that cellular responses depend on the timing of perturbation relative to the growth phase \cite{kakavand2025hydrodynamic}, we systematically varied the onset and duration of hydrodynamic forcing across multiple experimental rounds (Fig.~\ref{Fig_schematic_rounds}). For clarity, we grouped all treatments into "standard" scenarios, in which cultures were initially static and subsequently exposed to orbital shaking, and "reverse" scenarios, in which cultures experienced hydrodynamic forcing from inoculation and were later returned to static conditions.

In the 0-hour scenario, cultures were transferred to the orbital shaker immediately after inoculation and remained under hydrodynamic perturbation for the entire duration of the experiment. The shaker was placed inside the same incubator as the static controls and operated at 110~revolutions per minute (rpm), imposing a gentle, reproducible hydrodynamic motion. Light intensity, temperature, and nutrient background were identical between control and treatment groups; the presence or absence of hydrodynamic motion was the only experimental variable. Static control samples remained unshaken throughout and served as the reference condition.

In the 120-hour, 168-hour, and 348-hour delay scenarios, which collectively constitute the standard (static\,$\rightarrow$\,shaker) histories, cultures were kept static for 120, 168, or 348~h post-inoculation, respectively, before being moved to the orbital shaker for the remainder of the experiment. These onset times were chosen to coincide with the early exponential, mid-exponential, and stationary phases of the observed growth curves, thereby sampling distinct physiological stages along the population trajectory.

Reverse scenarios were designed to test whether the effects of early hydrodynamic perturbation are reversible once cultures are returned to quiescent conditions. In the 135-hour early perturbation scenario, cultures were exposed to orbital shaking immediately after inoculation, maintained under dynamic conditions for 135~h, and then transferred back to static conditions for the rest of the experiment. This finite-duration treatment allowed us to contrast continuous forcing (0-hour delay scenario) with time-limited early forcing at the same hydrodynamic intensity. In addition to the 135-hour lead perturbation scenario, we implemented a second reverse treatment in which cultures were transferred from shaker to static conditions at 168~h post-inoculation; both reverse scenarios are summarised schematically in Fig.~\ref{Fig_schematic_rounds}.

All standard and reverse scenarios were implemented for both \textit{H.~akashiwo} strains (HA452 and HA3107), each with matched static controls under identical environmental conditions. Unless otherwise noted, experiments were conducted with atleast two independent biological replicates per condition. The shaker speed (110~rpm) was held constant across all dynamic treatments to isolate the effect of onset timing and hydrodynamic cues history. The intensity of hydrodynamic forcing in the shaken cultures was quantified via the mean kinetic energy dissipation rate, as detailed later in this report.

\subsection*{Quantification of kinetic energy dissipation rate in dynamic culture}
\label{kinetic energy}

To characterise the intensity of hydrodynamic forcing applied in the dynamic treatments, we followed the approach of Gallardo Rodr\'{i}guez et al.~(2009) \cite{GallardoRodriguez2009}, which provides an estimate of the mean turbulent kinetic energy dissipation rate, $\epsilon$, in fluids agitated on an orbital shaker. This quantity represents the average rate at which kinetic energy is lost per unit mass of fluid due to viscous dissipation once the system has reached steady-state motion.

The energy dissipation rate was calculated using:
\begin{equation}
    \epsilon = \frac{1.94 \cdot n^3 \cdot D^4}{V_f^{2/3} \cdot \left(\frac{\rho_f \cdot n \cdot D^2}{\mu_f}\right)^{0.2}},
    \label{eq:epsilon}
\end{equation}
where $n$ is the orbital rotation speed (s$^{-1}$), $D$ is a characteristic length scale (m), and $V_f$ is the fluid volume (m$^{3}$). The fluid density $\rho_f$ and dynamic viscosity $\mu_f$ were taken as those of seawater with a salinity of 35~PSU at 20~$^\circ$C, approximately $1025~\mathrm{kg\,m^{-3}}$ and $1.1~\mathrm{mPa\,s}$, respectively \cite{Millero1981}. In our configuration, cultures were maintained in 55~mL glass tubes (ROTH ROTILABO) with an inner diameter of 23.0~mm and a working volume $V_f$ of 27~mL (25~mL fresh f/2--Si medium plus 2~mL inoculum), corresponding to a liquid column height of approximately 7.5~cm. The orbital shaker operated with an orbit diameter of 20~mm. These geometric parameters, together with the prescribed rotation speed (110 rpm), define the hydrodynamic regime experienced by the cells.

In practice, the shaken cultures were considered to have reached a steady hydrodynamic state when three conditions were met: (i) the platform rotation speed $n$ corresponding to 110~revolutions per minute (rpm) remained stable within $\pm 1\%$ of the setpoint according to the controller readout for at least 5~minutes; (ii) the swirling pattern in the culture tube, as visualised by the meniscus trajectory, became time-invariant over the same interval; and (iii) the incubator temperature stabilised at 22~$^\circ$C (within $\pm 0.2~^\circ$C), ensuring that $\rho_f$ and $\mu_f$ could be treated as constant. Under these conditions, $n$, $D$, $V_f$, $\rho_f$, and $\mu_f$ were taken as time-independent, and Eq.~\ref{eq:epsilon} was evaluated accordingly.

To verify the time required to attain this steady swirling regime, we performed control tests using a food dye solution with density matched to the algal cultures. The dye-containing tubes were placed on the orbital shaker under the same geometric and rotational settings, and the evolution of the colour field was visually monitored. These tests showed that the flow became homogeneously mixed and the meniscus trajectory stationary in less than one hour, a timescale negligible compared to the multi-week duration of the algal growth experiments. Thus, for the purposes of energy-dissipation estimation, $\epsilon$ was treated as a steady-state property over the course of the experiments.

Substituting the experimental parameters into Eq.~\ref{eq:epsilon} yielded an energy dissipation rate of $\epsilon = 9.54 \times 10^{-4}~\mathrm{W\,kg^{-1}}$ (i.e. $0.000954~\mathrm{m^{2}\,s^{-3}}$). This level of kinetic energy dissipation is approximately one order of magnitude higher than the upper bound reported for naturally occurring turbulence in the open ocean \cite{Sengupta2017Phytoplankton}, and thus represents a relatively strong but well-defined hydrodynamic perturbation.

Before settling on the standard operating condition, we carried out exploratory trials at three orbital speeds (50, 110, and 220~rpm) using the same tube geometry and working volume. At 50~rpm, the imposed motion was too weak to produce detectable differences from static controls in either growth or behaviour. At 220~rpm, by contrast, cells exhibited clear signs of mechanical stress, including a more rounded morphology and elevated susceptibility to lysis following Nile~Red staining to label cytoplasmic lipid droplets, indicative of excessive shear. The intermediate setting of 110~rpm provided a reproducible hydrodynamic perturbation that robustly distinguished dynamic cultures from static controls without inducing overt morphological or physiological damage. Throughout the two-week experimental window, phase-contrast microscopy confirmed that cells in the 0-hour delay scenario remained motile and morphologically intact under this condition, supporting the choice of 110~rpm for all systematic experiments.

\subsection*{Quantifying \textit{H. akashiwo} cell concentration and growth kinetics}
\label{subsec:growth curve}

Cell concentration and growth kinetics were quantified following the counting and analysis pipeline established in our previous work \cite{kakavand2025hydrodynamic}, Utilizing the same instrumentation and processing parameters, we summarize the procedure here and refer to \cite{kakavand2025hydrodynamic} for further technical details.

Small aliquots were withdrawn from each culture tube at predefined time points spanning lag, exponential, and stationary phases (sampling daily for the first 10~days, then at \(\sim\)24--48~h intervals thereafter; specific time points are given in Section~\ref{Impact of hydrodynamic perturbation timing on phytoplankton growth}). Before sampling, each tube was gently homogenised by slowly rolling it along its long axis two to three times, and the aliquot was taken from the mid-water column along the central axis, avoiding both the air–liquid interface and the bottom region. The same sampling protocol was applied to static and shaken cultures to ensure procedural consistency.

For counting, the well-mixed suspension was loaded into a custom PMMA (polymethyl methacrylate) chamber with internal dimensions \(3~\mathrm{mm} \times 2~\mathrm{mm} \times 0.56~\mathrm{mm}\), corresponding to a volume of \(\approx 3.4~\mu\mathrm{L}\) \cite{Sengupta2022Active}. The chamber was imaged on a Nikon\textsuperscript{\textregistered} SMZ1270 stereomicroscope. For each sample, short time-lapse sequences were acquired at 16~frames~s\(^{-1}\) for 10~s, with the field of view adjusted such that the entire chamber cross-section was captured at \(\sim 8\times\) camera zoom. For high-density samples (typically later than 120 h post-inoculation), cultures were diluted prior to imaging, and the final concentration was back-calculated using the known dilution factor.

Image sequences were processed using MATLAB Computer Vision Toolbox (Version~9.10.0. 1669831), as described in \cite{kakavand2025hydrodynamic}. Briefly, cells were detected and counted in each frame, and the counts were averaged over all frames in the 10~s video to give a single estimate per technical replicate. Concentrations were obtained by dividing the mean count by the chamber volume to yield cells~\(\mu\)L\(^{-1}\), and then scaled to cells~mL\(^{-1}\). For each time point and condition, we acquired four measurements (two independent biological replicates \(\times\) two technical replicates) and report the mean \(\pm\) s.d. across these replicates. The resulting concentrations were plotted against time to construct growth curves for each strain and hydrodynamic history.

Because the initial cell concentration is critical for interpreting subsequent growth, we standardised the inoculum across all tubes. Immediately before inoculation, the mother culture was gently mixed by slow axial rolling, and its concentration (cells~mL\(^{-1}\)) was measured using the same chamber-based method. Experimental cultures were then established by transferring 2~mL of this quantified mother culture into 25~mL of fresh f/2--Si medium (final volume 27~mL). The same mother culture and inoculum volume were used for all tubes within a given experimental round, with brief re-rolling between successive withdrawals to maintain homogeneity. Thus, the initial total cell number per tube (mother-culture concentration \(\times\) 2~mL) and the corresponding initial concentration (this total divided by 27~mL) were effectively matched across all treatments at \(t = 0\) without repeatedly sampling newly inoculated tubes.

To quantify population kinetics, we fitted a logistic growth model to the measured concentrations over time, following \cite{kakavand2025hydrodynamic}. The population size \(P(t)\) was described by
\begin{equation}
P(t) = \frac{K}{1 + \left(\dfrac{K - P_0}{P_0}\right)e^{-rt}},
\end{equation}
where \(K\) is the carrying capacity, \(P_0\) the initial population size, and \(r\) the specific growth rate. Parameter estimation (\(K\) and \(r\)) was performed by nonlinear regression of the logistic model to the data for each strain and treatment. The doubling time during exponential growth was then calculated as
\begin{equation}
T_d = \frac{\ln 2}{r}.
\end{equation}
Semi-logarithmic plots of the growth curves were used solely for visualising the exponential phase; all quantitative parameters were obtained from the logistic fits.

The image-based counting pipeline has been cross-validated against independent methods, including flow cytometry (chlorophyll autofluorescence gating) and high-magnification particle tracking, demonstrating agreement within approximately 10\% across a broad range of concentrations \cite{kakavand2025hydrodynamic}. This provides confidence that the derived growth rates and carrying capacities reflect the true population dynamics under both static and hydrodynamically perturbed conditions.

\subsection*{Quantitative analysis of single-cell swimming behavior}

Motility parameters were extracted by tracking individual cells in a custom-built imaging setup. Cells were introduced into a small observation chamber \((10\,\mathrm{mm}\times 3.3\,\mathrm{mm}\times 2\,\mathrm{mm})\), which was mounted on a custom-designed cage equipped with $x$--$z$ translation screws. The chamber was visualized using a lens--camera system and imaged at approximately \(1.7\times\) magnification with a USB-3 CMOS camera. The chamber was illuminated using a red LED light source (\SI{630}{\nano\metre} wavelength), and the entire chamber was captured within a single field of view. The complete experimental setup and computational method are described in detail in Ref.~\cite{Sengupta2022Active}.

After a 10-minute period, during which cells were allowed to reach a steady-state distribution within the millifluidic chamber following their introduction, the chamber was rotated 180$^\circ$ within 2 seconds using an Arduino-controlled stepper motor programmed in-house. As the cells swam to re-establish their steady-state distribution, short videos were recorded at 16~fps. From each recording, 120 frames (starting at the 16th frame to exclude post-rotation turbulence) were selected, covering cells movement.

Image processing was performed in Python using OpenCV2 to binarize and threshold images for cell identification. The Trackpy module was used to link cell positions across frames and generate full trajectories. From these trajectories, total swimming speed as well as horizontal and vertical components of velocity were calculated.

To ensure reliable data, only high-quality tracks were retained. The following exclusion criteria were applied: (a) track length shorter than 45 frames, (b) total displacement less than 2~$\mu$m (approximately one-tenth of a cell body length), and (c) swimming speeds below 10~$\mu$m/s or above 200~$\mu$m/s.

\subsection*{Estimating orientational stability from reorientation dynamics}

After a steady spatial distribution of cells inside the milifluidic chamber was reached, the observation chamber was rotated by $180^{\circ}$, as in our earlier work \cite{Sengupta2022Active}. Once cells had re-established upward swimming and equilibrated near the chamber mid-height, we performed three additional $180^{\circ}$ flips with pause intervals of 15, 12, and 12~s; the first two flips were analysed as separate sub-replicates. Videos were recorded at 16~fps; for each flip, we analysed 160 frames (10~s), which captures one reorientation event.

Cell trajectories were extracted using the same tracking pipeline described above. Tracks shorter than 45 frames or with net displacement $<\!20$~\textmu m were excluded; remaining tracks were smoothed with a quadratic (second-order) interpolation. For each track and frame, the instantaneous swimming orientation $\theta\in[-\pi,\pi]$ was computed from the smoothed displacements using a quadrant-resolved arctangent function, $\theta = \mathrm{atan2}(dy,dx)$; the coordinate convention follows \cite{chen2018}. Instantaneous angular velocity was then obtained as $\omega=\Delta\theta/\Delta t$.

To estimate the reorientation timescale, $\theta$ values were binned in 20$^{\circ}$ intervals from $-180^{\circ}$ to $+180^{\circ}$ and $\omega$ was averaged within each bin; outliers with $|\omega|>0.5$~rad\,s$^{-1}$ were discarded. The resulting $\omega(\theta)$ relationship was fitted to a sinusoid,
\begin{equation}
    \omega(\theta) \approx A \sin\theta
\end{equation}

and the reorientation timescale was defined as $\tau_r=1/(2A)$, following the low-Reynolds-number torque balance that links restoring torque to angular velocity \cite{roberts1970,mogami2001}. As discussed in \cite{Sengupta2022Active}, this $\tau_r$ represents a net orientational stability that integrates passive (hydromechanical) and any potential active reorientation contributions by the cells.

\subsection*{Estimation of lipid production}
\label{subsec:lipid_evaluation}

Lipid-droplet (LD) biosynthesis and accumulation were quantified by Nile Red staining coupled to single-cell image analysis, following the protocol and image-processing pipeline described in \cite{Sengupta2022Active,kakavand2025hydrodynamic}. At each predefined time point, cultures were first gently mixed by slow axial rolling to obtain a homogeneous cell suspension before sampling. Neutral lipids were stained with Nile Red (Thermo Fisher Scientific; excitation/emission 552/636~nm). A working staining mixture was prepared by combining 10~$\mu$L of 100~$\mu$M Nile Red in dimethyl sulfoxide (DMSO) with 200~$\mu$L of \textit{H.~akashiwo} culture supernatant and vortexing to ensure even dye dispersion; 200~$\mu$L of well-mixed cell suspension was then added, yielding a final dye concentration of 2.4~$\mu$M. Samples were incubated in the dark at room temperature for 15~minutes. This short incubation at low dye concentration was chosen to minimize dye exchange or leakage from LDs and to avoid detectable perturbation of cell physiology.

Cells were imaged live, without chemical fixation. After staining, a small droplet of the sample was placed on a glass slide, covered with a coverslip, and mounted on an Olympus CKX53 inverted microscope equipped with a DFK33UX265 camera. The slide–coverslip configuration imposed only mild mechanical confinement: cells remained viable and motile, but typically slowed after $\sim$15~minutes, which facilitated stable imaging. LD accumulation was evaluated using paired phase-contrast and fluorescence recordings. For each candidate cell, a phase-contrast image sequence was acquired first to confirm intact morphology (no lysis-like features) and active motility; only those cells that met these criteria were subsequently recorded in the fluorescence channel and retained for analysis. To limit phototoxic damage, the 552~nm excitation LED was operated at low output (instrument setting~5), and a TRITC-equivalent filter set with a narrow emission band centred near 630~nm and strong rejection above $\sim$660~nm was used to reduce chlorophyll autofluorescence background.

Time-series image stacks were recorded at 16~frames~s$^{-1}$ for 10~s. Across biological and technical replicates, several hundred cells were imaged per time point; for quantitative comparison across conditions, a random subset of at least 20 cells per biological replicate (with corresponding technical replicates) was selected for detailed analysis. Image sequences were processed using MATLAB (Image Processing Toolbox; MATLAB Computer Vision routines) and ImageJ, as in \cite{Sengupta2022Active}. After conversion to suitable grayscale representations, LDs were segmented by intensity-based thresholding to extract LD contours. From each LD, minimum and maximum Feret diameters were measured and used to estimate an effective LD volume under the assumption of a prolate-spheroid shape. The radius of an equivalent-volume sphere was then calculated and used to define a standardised LD cross-sectional area metric. The same segmentation and geometric workflow was applied to the cell body to obtain the cell cross-sectional area (see \cite{Sengupta2022Active} and \cite{kakavand2025hydrodynamic} for additional details).

For each analysed cell, the normalised lipid area was defined as the ratio of LD cross-sectional area to cell cross-sectional area. This single-cell neutral-lipid metric was then aggregated across cells, replicates, and treatments to characterise how hydrodynamic history modulates intracellular lipid accumulation. In the present study, we focused on this microscopy-based single-cell readout in combination with population growth and photophysiology; bulk biochemical lipid assays (e.g. fluorimetry, gravimetric extraction, or GC–FAME) were not performed here and are left for future work.

\subsection*{Flow cytometry}

Flow cytometry was used to validate and refine the image-based cell-counting procedure, ensuring accurate and reproducible estimates of cell concentration. Measurements were performed on an Attune\texttrademark~NxT Acoustic Focusing Cytometer (Thermo Fisher Scientific) operated in volumetric counting mode, which allows direct event-based quantification without the need for calibration beads. In this assay, the objective was to determine cell number rather than physiological status; accordingly, no additional stains were used.

For each condition, randomly selected samples from multiple time points were analysed. Aliquots with estimated concentrations exceeding 10{,}000 cells (typically after 100–120~h post-inoculation) were briefly diluted in Milli-Q\textregistered{} water to improve optical clarity and minimize coincidence artefacts and detector saturation. \textit{H.~akashiwo} cells were identified by intrinsic chlorophyll autofluorescence excited at 488~nm and detected in the BL3-H channel. A sequential gating strategy was applied: (i) fluorescence-based gating to isolate the chlorophyll-autofluorescent population, followed by (ii) forward- (FSC) and side-scatter (SSC) gating to exclude debris and doublets and to retain events with the expected size and internal complexity ~\cite{kakavand2025hydrodynamic}.

For each run, acquisition volume and any applied dilution factors were recorded, and final cell concentrations (cells~mL\textsuperscript{$-1$}) were computed by scaling the gated event counts accordingly. These flow-cytometric concentrations were then compared against values obtained from the image-based MATLAB counting pipeline, with discrepancies consistently below 10\%. This agreement confirms that the image-based method provides reliable estimates of cell concentration across the range of densities encountered in the static and hydrodynamically perturbed cultures.



\subsection*{Statistical analysis}

Paired t-tests were used to compare total lipid droplet (LD) volume, normalized LD area, and key photophysiological parameters between control and experimental groups at each sampled time point, covering both exponential and stationary growth phases. To evaluate differences in carrying capacity and doubling time between groups, two-sample t-tests were applied. A significance threshold of \(P = 0.05\) was used for all tests. Statistical analyses were conducted using GraphPad Prism software.

\newpage

\bibliography{paper.bib}


\bibliographystyle{unsrt}

\end{document}